\documentclass[12pt,preprint]{aastex}
\usepackage{epsfig}
\usepackage{amsmath}
\usepackage{color}
 \newcommand{\bm}[1]{\mbox{\boldmath$#1$}}

 \def\ed{\end{document}}
 \def\cb{\color{blue}}

\begin{document}
\title{Estimating Dark Matter
Distributions}

\author{Xiao Wang and Michael Woodroofe \affil{Department
  of Statistics, University of Michigan, Ann Arbor, MI
  48109-1092}\email{\tt wangxiao@umich.edu, michaelw@umich.edu}}
\author{Matthew G. Walker and Mario Mateo \affil{Department of
    Astronomy, University of Michigan, Ann
    Arbor, MI 48109-1090}\email{\tt mgwalker@umich.edu, mmateo@umich.edu}}
\and
\author{Edward Olszewski \affil{Steward Observatory, University of Arizona,
Tucson, AZ, 85721-0065}\email{\tt edo@as.arizona.edu}}

\begin{abstract}
  Thanks to instrumental advances, new, very large kinematic datasets for
  nearby dwarf spheroidal (dSph) galaxies are on the horizon.  A key
  aim of these datasets is to help determine the distribution of dark
  matter in these galaxies.  Past analyses have generally relied on
  specific dynamical models or highly restrictive dynamical
  assumptions.  We describe a new, non-parametric analysis of the
  kinematics of nearby dSph galaxies designed to take full advantage
  of the future large datasets.  The method takes as input the
  projected positions and radial velocities of stars known to be
  members of the galaxies, but does not use any parametric dynamical
  model, nor the assumption that the mass distribution follows that of
  the visible matter.  The problem of estimating the radial mass
  distribution, $M(r)$ (the mass interior to true radius $r$), is
  converted into a problem of estimating a regression function
  non-parametrically.  From the Jeans Equation we show that the
  unknown regression function is subject to fundamental shape
  restrictions which we exploit in our analysis using statistical
  techniques borrowed from isotonic estimation and spline smoothing.
  Simulations indicate that $M(r)$ can be estimated to within a factor
  of two or better with samples as small as 1000 stars over almost the
  entire radial range sampled by the kinematic data.  The technique is
  applied to a sample of 181 stars in the Fornax dSph galaxy.  We show
  that the galaxy contains a significant, extended dark halo some ten
  times more massive than its baryonic component.  Though applied here
  to dSph kinematics, this approach can be used in the analysis of any
  kinematically hot stellar system in which the radial velocity field
  is discretely sampled.
\end{abstract}

% \maketitle
% \author

 \section{Introduction}

 Despite their humble appearances, the dwarf spheroidal (dSph)
 satellites of the Milky Way provide a source of persistent
 intrigue.  Mysteries concerning their origin, evolution,
 mass density, and dynamical state make it difficult to know where
 to place these common galaxies in the context of standard (e.g.  Cold
 Dark Matter) models of structure formation.  Are they primordial
 building blocks of bigger galaxies, or debris from galaxy
 interactions?

 While dSph galaxies have stellar populations similar in number to
 those of globular clusters ($M_{Lum} \sim 10^{6-7} M_{\odot}$),
 their stars are spread over a much larger volume ($R \sim 2$-$6$ kpc
 compared to $10$-$50$ pc in globular clusters) resulting in the
 lowest luminosity (i.e., baryonic) densities known in any type of
 galaxy.  In many cases it is unclear how these galaxies could have
 avoided tidal disruption by the Milky Way over their lifetimes
 without the addition of considerable unseen mass.  This
 characteristic of dSph galaxies suggests that the dynamics of these
 systems are dominated either by significant amounts of unseen matter,
 or that these galaxies are all far from dynamical equilibrium.

 In general, the Jeans Equations ({\cb Equations (4-21), (4-24), and
   (4-27) of Binney \& Tremaine 1987 \cite{BT87}, hereafter, BT87})
 provide a robust description of the mass distribution, $M(r)$, of a
 collisionless gravitational system -- such as a dSph galaxy -- in
 viral equilibrium, Equation (\ref{eq:em}) below.  Their general form
 permits any number of mass components (stellar, gas, dark), as well
 as anisotropy in the velocity dispersion tensor and a non-spherical
 gravitational potential.  When applied to spherical stellar systems
 and assuming at most only radial or tangential velocity anisotropy,
 these equations can be simplified to estimate the radial mass
 distribution (Equation 4-55 of BT87):
 \begin{equation}
 \label{eq:mrjeans}
 M(r) = -\frac{r\overline{{v_r}^2}}{G}\biggl(\frac{\operatorname{d} \operatorname{ln} \nu}{\operatorname{d} \operatorname{ln} r}+
 \frac{\operatorname{d} \operatorname{ln} \overline{{v_r}^2}}{\operatorname{d} \operatorname{ln} r}+2\beta\biggr),
 \end{equation}
 where $\nu$ is the spatial density distribution of stars,
 $\overline{{v_r}^2}$
 is the mean squared stellar radial velocity at radius $r$.  The
 dimensionless isotropy parameter, $\beta(r)$, compares the system's radial and
 tangential velocity components:
 \begin{equation}
 \label{eq:jbeta}
 \beta\equiv 1-\frac{\overline{{v_\theta}^2}}{\overline{{v_r}^2}}.
 \end{equation}
 Apart from the constraints on the geometry and the functional form of the
 anisotropy, Equation (\ref{eq:mrjeans}) is model-independent, making
 it an appealing tool. It is relevant that Equation (\ref{eq:mrjeans}) and (\ref{eq:jns1}) below are applicable
 to any tracer population that in equilibrium and satisfies the collisionless Boltzman Equation.

 Kinematic datasets for individual dSph galaxies have historically
 been far too small (typically containing radial velocities for $\sim$
 30 stars; see Mateo 1998) to allow for a precise determination of
 $M(r)$ using relations similar to Equation (\ref{eq:mrjeans}).
 Instead, authors have been forced to adopt additional strong
 assumptions that reduce the Jeans Equation to even simpler forms and
 where the relevant distributions ($\nu(r)$ and $\overline{v_r^2}(r)$
 in Equation \ref{eq:mrjeans}) are represented by parametric models.
 Specifically, if one assumes isotropy of the velocity dispersion
 tensor (i.e., $\beta=0$), spherical symmetry, and that the starlight
 traces the mass distribution (effectively a single-component King
 model (Irwin and Hatzidimitriou 1995)), then one obtains for the M/L
 ratio (Richstone and Tremaine 1986):
 \begin{equation}
 \label{eq:moverl}
{{M}\over{L}} = \eta{{9\sigma^2_0}\over{2 \pi G I_0 R_h}},
\end{equation}
where $\sigma_0$ is the one-dimensional central velocity dispersion,
$I_0$ is the central surface brightness, and $R_h$ is the half-light
radius.  The parameter $\eta$ is nearly equal to unity for a wide
range of realistic spherical dynamical models so long as the mass
distribution is assumed to match that of the visible matter.  With
this approach -- the modern variant of the classical `King fitting'
procedure (King 1966) -- the measured central radial velocity
dispersion and surface brightness yield estimates of such quantities
as the global and central M/L ratios.  In all eight of the MW's
measured dSphs{\cb \footnote{We exclude the Sagittarius dSph, which is
    unambiguously undergoing tidal destruction (Majewski et al.
    2003).}}, large central velocity dispersions have conspired with
their low surface brightnesses to produce large inferred M/L values.
This line of reasoning has led to a general belief that dSph galaxies
are almost completely dark-matter dominated, and their halos have
assumed the role of the smallest non-baryonic mass concentrations
identified so far in the present-day Universe.

This analysis fails for galaxies that are far from dynamical equilibrium, for example due to the effects of
external tidal forces from the Milky Way (Fleck and Kuhn 2003; Klessen and Kroupa, 1998). Numerical models aimed
to investigate this (Oh et al.  1995; Piatek and Pryor 1995) generally found that tides have negligible effects
on the central dynamics of dSph galaxies until the survival time of the galaxy as a bound system becomes
comparable to the orbital time (about 1 Gyr for the closer dSph satellites of the Milky Way).  Observations
agree with this broad conclusion by finding that remote dSph galaxies are no less likely to contain significant
dark matter halos than systems located closer to their parent galaxy (Mateo et al. 1998; Vogt et al. 1995).
However, so-called resonance models (Fleck and Kuhn 2003; Kuhn 1993; Kuhn et al. 1996) have been proposed that
imply the central velocity dispersions can be significantly altered due to the inclusion of stars streaming
outward from the barycenter of a galaxy and projected near the galaxy core.  Recent versions of these models
invariably imply a significant extension of the affected galaxies along the line-of-sight (more precisely, along
the line between the center of the dwarf and the Milky Way; Kroupa 1997; Klessen and Kroupa 1998) and a massive
tidal stream along the satellite's orbit. Observations do not reveal strong evidence of significant
line-of-sight distortions in dSph galaxies (Hurley-Keller et al 1999; Klessen et al. 2003), other than
Sagittarius (e.g. Ibata et al. 1995); thus, for the purposes of this paper, we will assume that dSph galaxies
are generally close to a state of dynamical equilibrium.

Even with this enabling assumption, the classical analysis of dSph
masses as we describe it above is far from ideal for a number of
reasons.  First, though recent work (e.g. Irwin and Hatzidimitriou
1995) has helped to greatly improve estimates of dSph structural
parameters ($I_{0}$ and $R_{h}$ in Equation \ref{eq:moverl}), the
errors in the velocity dispersions -- often dominated by Poisson
uncertainties due to the small number of kinematic tracers --
contribute the principle source of uncertainty in M/L estimates.
Second, there is little reason to suppose the assumption that mass
follows light in dSphs is valid.  In all other galaxies the bulk of
the matter resides in a dark halo extending far beyond the luminous
matter, a trend that becomes more exaggerated toward smaller scales
(Kormendy and Freeman 2004).  Finally, velocity anisotropies, if they exist,
may mimic the presence of dark matter, and so represent a tricky
degeneracy in the model even if the assumption of isotropy is dropped.

Modern instrumentation is poised to deliver dramatically larger
kinematic datasets to help minimize the first problem.  For example,
the Michigan/Magellan Fiber System, now operational at the Magellan
6.5-m telescopes, obtains spectra from which high-precision radial
velocities can be measured of up to 256 objects simultaneously.  As a
result, it is now feasible to obtain thousands of individual stellar
spectra in many dSph systems, enlarging sample sizes by more than an
order of magnitude.  This not only reduces the statistical uncertainty
of the dispersion estimation, but can also provide information on the
spatial variation of the dispersion across the face of a galaxy.
These rich datasets therefore allow for fundamentally improved results
even using fairly conventional analysis techniques.  For example, by
parameterizing the velocity anisotropy, Wilkinson et al. (2002) and
Kleyna et al. (2002) show that samples of $\sim 200$ stars can begin
to break the degeneracy between anisotropy and mass in spherical
systems.

But these large datasets also allow us to aim higher.  In this paper,
we introduce and develop the formalism for a qualitatively different
sort of analysis designed to make the most efficient use of large
kinematic datasets.  Rather than adopting a model that parameterizes
the various distributions used in the Jeans equation (e.g., $M(r)$,
$\nu(r)$, $\overline{v_r^2}$, or $\beta(r)$), we operate on the star
count and radial velocity data directly to estimate the mass
distribution non-parametrically.  We estimate the true
three-dimensional mass distribution from the projected stellar
distribution and the line-of-sight velocity distribution.  In this
first application of our technique, we still require the assumptions
of viral equilibrium, spherical symmetry, and velocity isotropy
($\beta=0$).  In Section 2, we introduce notation and definitions of
the three-dimensional as well as projected stellar and phase-space
densities and review the Jeans Equation.  In Sections 3 and 4, we show
how to use very general shape constraints on the mass distribution to
estimate the detailed form of the velocity dispersion profile and the
radial mass distribution. We illustrate the process on simulated data,
demonstrating that, when furnished with large datasets, nonparametric
analysis is a powerful and robust tool for estimating mass
distributions in spherical or near-spherical systems (Section 5).  We
illustrate the application of our approach to an existing -- but
relatively small -- kinematic dataset for the Fornax dSph galaxy at the
end of Section 5.

Our emphasis in this paper is the application of our technique to dSph kinematics, but the method described in
this paper is applicable to any dynamically hot system in which the radial velocity field is sampled discretely
at the positions of a tracer population.  Thus, our methodology would work for, say, samples of globular
clusters or planetary nebulae surrounding large elliptical galaxies, or for individual stars within a globular
cluster.  Our approach follows earlier non-parametric analysis of globular cluster kinematics by Gebhardt and
Fischer (1995) and Merritt et al. (1997), though the details of our method differ significantly.

\section{Jeans' Equation}

Let $\bm{X} = (X_1,X_2,X_3)$ and $\bm{V} = (V_1,V_2,V_3)$
denote the 3-dimensional position and velocity of a star within a
galaxy.  We will
regard these as jointly distributed random vectors, as in BT87, p.
194.  Suppose that ${\bm X}$ and ${\bm V}$ have a joint density that
is spherically symmetric and isotropic, so that
 \begin{equation}
 \label{eq:density}
         P[{\bm x} \le {\bm X} \le {\bm x} + d{\bm x}\ \&\ {\bm v} \le {\bm
   V} \le {\bm v}+d{\bm v}] = f_0(r,v)d{\bm x}d{\bm v}, \end{equation}
 where $r^2 = x_1^{2}+x_2^{2}+x_3^{2}$ and $v^2 =
 v_1^{2}+v_2^{2}+v_3^{2}$.  We also suppose that $\bm{V}$ has been centered
 to have a mean of zero so that $\int {\bm v} f_0(r,v)d{\bm
 x}d{\bm v} = 0.$ Suppose finally that the mass density $\rho(r)$ is
 spherically symmetric and let
 \begin{equation}
 \label{eq:em}
        M(r) = 4\pi \int_{0}^{r}
 t^2\rho(t)dt,
 \end{equation}
 the total mass within $r$ of the center.  The
 goal of this paper is to come up with a means of estimating $M(r)$
 non-parametrically, specifically without assuming any special
 functional form for $f_0(r,v)$ or $\rho(r)$.  In the presence of spherical
 symmetry, the relevant form of the Jeans Equation is
 \begin{equation}
 \label{eq:jns1}
         M(r) = -{r^2\mu(r)\over G}{d\over dr}\log\big[f(r)\mu(r)\big],
 \end{equation}
 where
 \begin{equation}
 \label{eq:mu}
         \mu(r) = {1\over 3 f(r)}\int v^2 f_0(r,v)d\bm{v}
 \end{equation}
 and $$ f(r) = \int f_0(r,v)d\bm{v}.  $$
 In statistical terminology, $\mu(r)$ is the conditional expectation
 of $V^2 = (V_1^{2}+V_2^{2}+V_3^{2})/3$ given ${\bm x}$ (this is the
 same as the conditional expectation on $V_3^2$, hence the factor of
 3), and the marginal density of $\bm{X}$ is $f(r)$.  In astronomical
 terms, $\mu(r)$ is the one-dimensional radial-velocity dispersion
 profile squared, and $f(r)$ is the true, three-dimensional density
 profile of the tracer population.  Using the notation of BT87, these correspond
 to $\overline{v_r^2}$ and $\nu(r)$, respectively.

 To estimate $M(r)$ from equation (\ref{eq:jns1}) clearly requires estimates of
 the functions $f$ and $\mu$.  For dSph galaxies, the data available
 for this consist of a large sample of positions and photometry of
 stars in individual systems (Irwin and Hatzidimitriou 1995; hereafter
 IH95), and much smaller samples of stars with positions and
 velocities (e.g.  Walker et al.  2004).  Of course, it is not
 possible yet to determine complete three-dimensional positional or
 velocity information for any of these stars, but only the velocity in
 the line of sight and the projection of position on the plane
 orthogonal to the line of sight.  With a proper choice of coordinates,
 these observables become $X_1,X_2$, and $V_3$.  These can be made
 equivalent to, say, right ascension and declination, and radial
 velocity, respectively.  The incomplete observation of $\bm{V}$ does
 not cause a problem here, given the assumed isotropy, since $v^2$ can
 be replaced by $3v_3^{2}$ in Equation (\ref{eq:mu}) without changing the value of $\mu(r)$.  The incomplete observation of position poses a more serious problem that is known as Wicksell's (1925) Problem in the
 statistical literature.  The procedures for estimating $f$ and $\mu$
 are consequently different; they will be considered separately in Sections 3 and 4.

 \section{The Spatial Distribution of Stars: Estimating $\bm{f}$}
 \label{sect:h}

 Let $R = \sqrt{X_1^{2}+X_2^{2}+X_3^{2}}$ and $S =
 \sqrt{X_1^{2}+X_2^{2}}$ denote the true three-dimensional distance of
 a star and its two-dimensional projected distance from the origin,
 respectively.  We denote the corresponding densities as $f_{\bf R}$
 and $g_{\bf S}$.  Then $f_{\bf R}(r) = 4\pi r^2f(r)$, and
 \begin{equation} \label{eq:fg}
         g_{\bf S}(s) = 4\pi s\int_{s}^{\infty} {f(r)rdr\over \sqrt{(r^2-s^2)}}.
 \end{equation}
 For an example, consider Plummer's distribution,
 \begin{equation}
 \label{eq:plummer}
        f_0(r,v) = {c_0\over b^5}\left[{b\over\sqrt{1+{1\over 3}r^2}}-
   {1\over 2}v^2\right]_{+}^{7\over 2},
\end{equation}
 where $c_0$ is a constant,and $b$ is a parameter that is related to the velocity dispersion through
$E(V^2\vert R=r) = {b/ [2\sqrt{1+r^2/3}]}$. Here $r$ is measured in units of 100 pc and $v$ in km/s. We also
employ here the notation $[x]_{+}$ to denote taking the larger of $x$ and $0$.  With
 this definition, we take $[x]_+^\alpha$ to be shorthand for
 $([x]_+)^\alpha$; that is, the value of $x$ is either positive or
 zero before raising to the power $\alpha$. For this case, $f_{\bf
   R}(r) = r^2/[\sqrt{3}\sqrt{(1+r^2/3)^5}]$ and $g_{\bf S}(s) =
 2s/[3(1+s^2/3)^2]$. We use the Plummer distribution in our
 simulations below.

        IH95 provide data from which $f$ can be estimated.  These data
 consist of a sample of stars with projected radii $S_1,\cdots,S_N$ and
 counts $N_k$ of the number of stars for which $r_{k-1} < S \le
 r_{k}$ where $0 = r_0 < r_1 < \cdots < r_m$ divide the stars into
 bins.  Let $G_{\bf S}$ denote the distribution function of $S$, so
 that $G_{\bf S}(s) = P[S \le s] = \int_{0}^{s} g_{\bf S}(t)dt$.  Then
 $G_{\bf S}(r_{k})$ may be estimated from its empirical version
 \begin{equation}
 \label{eq:estgee}
         G_{\bf S}^{\#}(r_{k}) = {N_1+\cdots,+N_k\over N}.
 \end{equation} In this context and throughout this paper, the $\#$
 symbol denotes functions estimated directly from empirical data.
 Since the samples of stars used by
 IH95 to determine the surface density of most of the Milky Way dSph
 galaxies are quite large, there is generally very
 little statistical uncertainty in our estimate of $G^\#_{\bf S}$ from
 Equation (\ref{eq:estgee}).  If we now model $f$ by a step function,
 say $f(r) = f_k$ for $r_{k-1} < r \le r_k$ and $k = 1,\cdots,m$, then
 $f_1,\cdots,f_m$ may be recovered directly.  In this case, the
 integral in Equation (\ref{eq:fg}) is easily computed leading to $$
         1-G_{\bf S}(r_{j-1}) = \sum_{k=j}^m a_{jk}f_k,
 $$
 where
 $$
        a_{jk} = {4\pi\over 3}\big[\sqrt{(r_{k}^2-r_{j-1}^2)^3} -
 \sqrt{(r_{k-1}^2-r_{j-1}^2)^3}\big].
 $$
 Then $f_1,\cdots,f_m$ may be recovered from
 \begin{equation}
 \label{eq:gf}
         f_j = {1\over a_{jj}}\big[1-G_{\bf S}(r_{j-1}) - \sum_{k=j+1}^m a_{jk} f_k\big].
 \end{equation}
 The summation is to be interpreted to be zero when $j = m$, and the
 right side of equation (\ref{eq:gf}) can in practice be estimated by
 replacing $G_{\bf S}$ with its empirical version $G_{\bf S}^{\#}$.

\begin{figure}
 \begin{center}
   \resizebox{3in}{3in}{\includegraphics{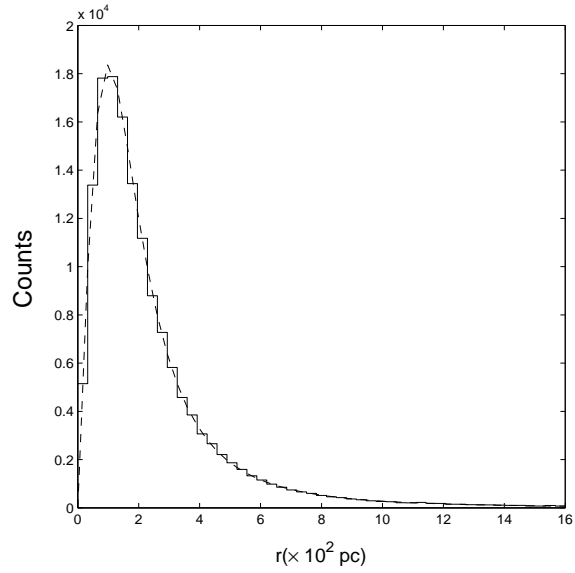}}
 \end{center}\caption{\label{fig:counts} \small Simulated projected
 counts for 150,000 stars drawn from a Plummer model.  The histogram
 shows the distribution of selected stars from this model; the dotted
 line shows the actual distribution defined by the model.}
 \end{figure}

 \begin{figure}
 \begin{center}
   \resizebox{3in}{3in}{\includegraphics{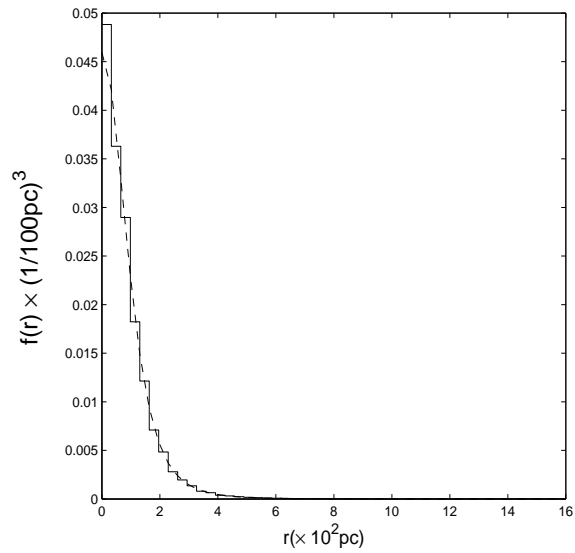}}
 \end{center}\caption{\label{fig:fbar} \small The function $f$ derived from
   the counts shown in Figure \ref{fig:counts} derived using Equation
   (\ref{eq:gf}) and for a sample size of 150,000 positions derived
   from a Plummer model.  The true distribution is shown as a dashed
   line.}
 \end{figure}

 The results of applying this method to a sample of 150,000 positions
 drawn from a Plummer model with $b=200\times 3.1\times 10^{15} km^3/sec^2$ are shown in Figures \ref{fig:counts} and
 \ref{fig:fbar}. Figure \ref{fig:counts} shows simulated counts of the number
 of stars in equally-spaced bins ranging from $0=r_0\le r_1\le
 \cdots\le r_{50}=16\times 100pc$ . The estimate of
 $f$ from these simulated data is shown in Figure \ref{fig:fbar}.  In
 both figures the dotted line denotes the true function.

 The estimation of $f$ follows Wicksell's (1925) original analysis and
 can almost certainly be improved.  We have not done that here,
 because the sample sizes are so large, but related questions are
 currently under investigation by Pal (2004).

 \section{Velocity Dispersion Profile: Estimating $\bm{\mu}$ and $M$}
 \label{sect:mu}

 For estimating $\mu$ and $M$ it is convenient to work with squared
 radius $Y = S^2 = X_1^{2}+X_2^{2}$, as in Groeneboon and Jongbloed
 (1995) and Hall and Smith (1992).  The joint density of $(Y,\bm{V})$
 is then
 \begin{equation}
 \label{eq:vy}
        g_{Y,\bf V}(y,{\bm v}) = \pi\int_{y}^{\infty} {f_0(\sqrt{z},v)
   \over\sqrt{z-y}}dz.
 \end{equation}
 Relation (7) represents a special case on integrating over all three
 velocity components and using $g_{\bf{S}}(s) = 2sg_{\bf{Y}}(s^2)$.  Now let
$$
        \phi(z) = \int v_3^{2}f_0(\sqrt{z},v)\bf{dv},
$$
and
$$
        \psi(y) = \int v_3^{2}g_{Y,\bf{V}}(y,\bf{v})d\bm{v},
$$
so that $\psi(y)/g_Y(y)$ is the conditional expectation of
$V_3^{2}$ given $Y = y$, and similarly for $\phi$.
 We also define the important function
 $$\Psi(y) = \int_{y}^{\infty} {\psi(t)dt\over\sqrt{t-y}}.  $$
 Then, reversing the orders of integration and recognizing a
 beta integral (see BT87, p. 205), we can write
 \begin{equation}
 \label{eq:Psi}
         \Psi(y) = \pi^2\int_{y}^{\infty} \phi(z)dz
 \end{equation}
 and
 \begin{equation}
 \label{eq:psinv}
         \phi(z) = -{1\over\pi^2}\Psi'(z),
 \end{equation}
 where the prime notation ($'$) denotes the derivative.
 Clearly, $f(r)\mu(r) = \phi(r^2)$ in Equation (\ref{eq:mu}).
 Thus, the expression for $M(r)$ from the Jeans equation (Equation (\ref{eq:jns1})) can
 be rewritten as
 \begin{equation}
 \label{eq:jns2}
         M(r) = {2r^3\over G\pi^2}{\Psi''(r^2)\over f(r)}.
 \end{equation}

 Our estimator for $f$ was obtained in Section 3,
 so we focus here on determining $\Psi''$.  For our problem, we benefit
 from noting that the function $\Psi$ is subject to certain shape
 restrictions which are useful in the estimation process.  From
 Equation (\ref{eq:Psi}) we can see that $\Psi$ is a non-negative
 decreasing function, while from Equation (\ref{eq:jns2}) it is evident that
 $\Psi$ has a non-negative second derivative (that is, it is upward
 convex).  In fact, $r^3\Psi''(r^2)/f(r)$ must be an increasing
 function that vanishes when $r = 0$, physically consistent with its
 direct proportionality to $M(r)$ in Equation (\ref{eq:jns2}).

 The next step in the estimation of $M(r)$ is to estimate $\Psi$.
 Suppose that there is a sample $(X_{i,1},X_{i,2},U_i),\ i =
 1,\cdots,n$ of projected positions $(X_{i,1},X_{i,2})$ and radial
 velocities $U_i$ measured with known error, $\epsilon_i$.  Thus, $U_i
 = V_{i,3}+\epsilon_i$, where where $\epsilon_1,\cdots,\epsilon_n$ are
 independent of $(X_{i,1},X_{i,2},V_{i,3}),\ i = 1,\cdots,n$ with zero
 mean values and known variances, $\sigma_1^{2},\cdots, \sigma_n^{2}$.
 In practice, there may be selection effects: An astronomer may choose to
 sample some regions of a galaxy more intensely than others, or
 external factors (weather, moonlight, telescope/instrument problems)
 may cause undersampling of some regions.  To address this, we must
 include a selection function, $w_0(x_1,x_2)$, into the model.  Here,
 $w_0(x_1,x_2)$ is the probability that a star in a galaxy is included
 in the kinematic sample, given that it is there. Thus the selection
 probability is assumed to depend on projected position, $(x_1,x_2)$,
 only.  As before, let $Y_i = X_{i,1}^{2}+X_{i,2}^{2}$; then the joint
 density of $Y_i$ and ${\bm V}_i$ in the sample is
\begin{equation}
\label{eq:slctn1}
        g_{{\bf Y,V}}^*(y,{\bm v}) = w(y)g_{{\bf Y,V}}(y,{\bm v}),
\end{equation}
where
$$
        w(y) = {1\over c}\int_{-\pi}^{\pi} w_0[\sqrt{y}\cos(\theta),\sqrt{y}\sin(\theta)] d\theta,
        $$
        $g_{{\bf Y,V}}$ is as in Equation (\ref{eq:vy} ), and $c$
        is a normalizing constant, determined by the condition that
        the integral of $g_{{\bf Y,V}}^*(y,{\bm v})$ be one.
        Integrating over ${\bm v}$ in (\ref{eq:slctn1}), the actual
        density of $Y$ is
\begin{equation}
\label{eq:slctn2}
        g_{{\bf Y}}^*(y) = w(y)g_{{\bf Y}}(y),
\end{equation}
and $c$ is determined by the condition that $\int g_{\bf Y}^*(y)dy =
1$.  Here $g_{\bf Y}$ may be determined from the complete
photometric data for a given galaxy to give the projected stellar
distribution as a function of $Y$. If
$f$ is a step function, then combining equations (\ref{eq:fg}) and
(\ref{eq:gf}) with the relation $g_{\bf S}(s) = 2sg_{\bf Y}(s^2)$
leads to
$$
        g_{\bf Y}(y) = 2\pi\left[f_j\sqrt{r_j^{2}-y} + \sum_{k=j+1}^m f_k \left(\sqrt{r_k^{2}-y} - \sqrt{r_{k-1}^{2}-y}\right)\right]
        $$
        in the interval $r_{j-1}^2 \le y \le r_j^{2}$ for $j =
        1,\cdots,m$.  If an astronomer can specify $w_0(x_1,x_2)$,
        then $c$ and $w(y)$ can be computed directly using Equation
        (15).  Then
 \begin{equation}
 \label{eq:Psisharp}
   \Psi^{\#}(z) = {1\over n}\sum_{i:Y_i>z} {U_i^{2}-\sigma_i^{2}\over
  w(Y_i) \sqrt{Y_i-z}}
 \end{equation}
 is an unbiased estimator of $\Psi(z)$ for each $z$;
 that is, $\langle \Psi^\#(z)\rangle = \Psi(z)$ for each $z$.

If an astronomer cannot specify $w_0$, then it must be estimated from
the data.  The first step is to estimate $g_{\bf Y}^*$, which can in
principle be done in many ways.  The simplest is to use a kernel
estimator of the form
$$
        \hat{g}_{\bf Y}^*(y) = {1\over nb}\sum_{i=1}^n K\left({Y_i-y\over b}\right),
        $$
        where $K$ is a probability density function, called the
        {\it kernel}, and $b$ is a positive factor called the {\it
          bandwidth} that may be specified by the user or computed
        from the data.  Silverman (1986) is a recommended source for
        background information on kernel density estimation.  Possible
        choices of $K$ and the computation of $b$ are
        discussed there.  Other methods for estimating $g_{\bf Y}^*$ include
        local polynomials [79-82, Loader 1999] and log-splines [178-179, Gu 2002].  Whichever method is adopted, once $g_{\bf
          Y}^*$ has been estimated, $w$ may be subsequently estimated by
\begin{equation}
\label{eq:sltnest}
        \hat{w}(y) = {\hat{g}_{\bf Y}^*(y)\over g_{\bf Y}(y)}.
\end{equation}
Then $\Psi$ can be estimated by replacing $w$ with $\hat{w}$ in (\ref{eq:Psisharp}). In this case $\Psi^{\#}$ is
no longer exactly unbiased, but it is at least consistent in the statistical
sense that as the number of data points increases, the estimator
continuously tends ever closer to the true value.

Generally, $\Psi^{\#}(z)$ does not satisfy the shape restrictions when
viewed as a function of $z$: $\Psi^{\#}(z)$ is unbounded as $z$
approaches any of the data points $Y_i$.  It is decidedly not monotone
nor convex (see Figure \ref{fig:Psi}).  For these reasons $\Psi^{\#}$
may be called the {\it naive} estimator.  This naive estimator can be
improved by imposing physically justified shape restrictions on
$M(r)$.  One can imagine a wide range of possible model forms for
$M(r)$, but in the spirit of attempting a non-parametric solution, we
adopt a simple spline of the form
 \begin{equation}
 \label{eq:spln1}
         M(r) = \sum_{i=1}^m \beta_i[r-r_{i-1}]^p_{+},
 \end{equation}
 where $0 = r_0 < r_1 < \cdots < r_m$ are knots,
 $\beta_1,\cdots,\beta_m$ are constants, and $x_{+}^{p} = [\max(0,x)]^p$.  The simplest case is for $p
 = 1$, but this form for $M(r)$ leads to a divergent estimate of the
 central velocity dispersion, $\mu(0)$.  For $p = 3$, $M(r) \sim
 \beta_1r^{3}$ as $r \to 0$; thus, $3 \beta_1/4\pi$ provides an
 estimate of the central mass density with the correct asymptotic
 behavior at the galaxy center.  Unfortunately, translating the shape
 restrictions on $M(r)$ into conditions on $\beta_1,\cdots,\beta_m$ is
 much more difficult when $p = 3$ than for $p \leq 2$.  For this paper
 we have chosen to compromise with $p = 2$ because it leads to a good
 estimate of $\mu(0)$ yet is still amenable to the application of the
 shape restrictions we described above.  We plan to tackle the case $p
 = 3$ in a future paper.

For our adopted case ($p = 2$) the expression (Equation
(\ref{eq:spln1})) for the mass, $M(r)$, is a quadratic spline:
 \begin{equation}
 \label{eq:spln2}
         M(r) = \sum_{i=1}^m \beta_i[r-r_{i-1}]^2_{+}.
 \end{equation}
 One natural constraint is that $M(r)$
 remain bounded as $r \to \infty$. Expanding Equation (\ref{eq:spln2})
 to read
 \begin{equation}
 \label{eq:mr2}
  M(r) = \left(\sum_{i=1}^{m} \beta_i\right)r^2 - 2\left(\sum_{i=1}^{m}
   \beta_i r_{i-1}\right)r + \sum_{i=1}^{m} \beta_i r_{i-1}^2,
 \end{equation}
 for $r > r_{m-1}$,
 we see that this constraint is equivalent to requiring that
 $\sum_{i=1}^m \beta_i = 0$, and $\sum_{i=1}^m \beta_i r_{i-1} = 0$.
 From this first equality constraint ($\sum_{i=1}^m \beta_i = 0$),
we can trivially say $\beta_m = -\sum_{i=1}^{m-1} \beta_i$ and, therefore,
\begin{equation}
 \label{eq:spln3}
         M(r) = \sum_{i=1}^{m-1} \beta_i\big\{[r-r_{i-1}]^2_{+} - [r-r_{m-1}]_{+}^{2}\big\}.
\end{equation}
We also require that $M(r)$ be a non-decreasing function of $r$ (no
negative mass), or
equivalently that the derivative $M'(r)$ be non-negative.  It is clear
from Equations (\ref{eq:spln2}) and (\ref{eq:spln3}) that $M'(r)$ is a
piecewise linear function that is constant on the interval
$r_{m-1} \le r \le r_m$.  Thus, the condition that $M(r)$ be
non-decreasing is equivalent to requiring that $M'(r_{i-1})\geq 0$ for
$i = 1, \cdots,m-1$.  That is,
 \begin{equation}
 \label{eq:cnstrnt1}
         \sum_{i=1}^j \beta_i (r_j - r_{i-1}) \ge 0
 \end{equation}
 for $j = 1,\cdots,m-1$. For the case $j = m - 1$, this constraint can
 be written as $\sum_{i=1}^{m-1} \beta_i (r_{m-1} - r_{i-1}) = 0$.
 This is clear from noting that we do not change this summation if we
 replace $m-1$ with $m$; thus we can write $\sum_{i=1}^{m-1} \beta_i
 (r_{m-1} - r_{i-1}) = \sum_{i=1}^m \beta_i (r_{m-1} - r_{i-1})$,
 where we have used the facts that $(\sum_{i=1}^m \beta_i) r_{m-1} = 0$
 and that $\sum_{i=1}^m \beta_i r_{i-1} = 0$ as required by Equation
 (\ref{eq:spln2}).  These are the constraints imposed on
 $\beta_1,\cdots,\beta_{m-1}$ that we employ below to estimate $M(r)$.

 If we solve for $\Psi''$ in Equation (\ref{eq:jns2}), we can write
      $$
        \Psi''(r^2) = {G\pi^2f(r)M(r)\over 2r^3} = \sum_{i=1}^{m-1}
        \beta_i\gamma_i(r^2),
 $$
 where
 $$
 \gamma_i(r^2) = {G\pi^2f(r)\over
   2r^3}\big([r-r_{i-1}]_{+}^{2}-[r-r_{m-1}]_{+}^{2}\big),
 $$
 For notational convenience, we
 define $y = r^2$, then solve for $\Psi(r^2) \equiv
 \Psi(y)$ in terms of the coefficients $\beta_i$ to get
 \begin{equation}
 \label{eq:Psi2}
         \Psi(y) = \int_{y}^{\infty} (t-y)\Psi''(t)dt = \sum_{i=1}^{m-1} \beta_i\Gamma_i(y),
 \end{equation}
 where
 $$
         \Gamma_i(y) = \int_{y}^{\infty} (t-y)\gamma_{i}(t)dt
 $$
 for $i = 1,\cdots,m-1$.

 \par

 The next step is to impose the shape restrictions implicit in
 Equation (\ref{eq:jns2}).  If $\Psi^{\#}$ were square integrable, this could
 be directly accomplished by minimizing the integral of
 $[\Psi^{\#}-\Psi]^2$, or equivalently by minimizing
 \begin{equation} \label{eq:crtrn1}
         \kappa = {1\over 2}\int_{0}^{\infty} \Psi(t)^2dt -
 \int_{0}^{\infty} \Psi^{\#}(t)\Psi(t)dt
 \end{equation}
 with respect to the coefficients, $\beta_i$.  The function
 $\Psi^{\#}$ is not integrable, essentially because $\int_{0}^{1} dx/x
 = \infty$, but $\kappa$ can still be minimized.  Letting ${\bf \beta}$ be
 the vector ${\bf \beta} = [\beta_1,\cdots,\beta_{m-1}]^T$, this criterion
 may be expressed in matrix notation as
 \begin{equation}
 \label{eq:kappa}
         \kappa = {1\over 2}{\bf \beta}^T Q {\bf\beta} - {\bf\beta}^T {\bf z},
 \end{equation}
 where the elements of the vector ${\bf z}$ are given by
 $$
         z_i = \int_{0}^{\infty} \Psi^{\#}(t)\Gamma_i(t)dt,
 $$
 and the elements of $Q$ are given by
 $$
 q_{ij} = \int_{0}^{\infty} \Gamma_i(t)\Gamma_j(t)dt $$
 for $i,j =
 1,\cdots,m$.  Thus, the estimation problem for $\Psi$ leads to a
 quadratic programming problem of minimizing $\kappa$ in Equation
 (\ref{eq:kappa}) subject to condition (\ref{eq:cnstrnt1}) with equality
 when $j = m-1$.

 If we can determine $z_i$ and $q_{ij}$, we can estimate the
 coefficients $\beta_i$ which in turn gives us our estimate
 for $M(r)$.
 Observe that the $z_i$ depend explicitly on both the deduced stellar
 density distribution, $f$, and the velocity dispersion profile
 derived from a set of radial velocity tracers, $\mu$.  The $q_{ij}$
 depend only on $f(r)$.  Let
 $\Psi^\#_2$ be defined as
\begin{eqnarray}
\label{eq:psin2}
\Psi_2^{\#}(z) &=& \int_0^{z}(z-y)\Psi^{\#}(y)dy   \nonumber\\
&=& {1\over n}\sum_{i: Y_i>z}{U_i^2-\sigma_i^2\over w(Y_i)}\left[{4\over
    3}\left((Y_i-z)^{3\over 2} - Y_i^{3\over 2}\right)
+2zY_i^{1\over 2}\right] \nonumber \\
&&\ \ \ \ \ \ \ \ \ \ +{1\over n}\sum_{i: Y_i\le z}{U_i^2-\sigma_i^2\over w(Y_i)}\left[-{4\over
    3} Y_i^{3\over 2}+2zY_i^{1\over 2} \right],
\end{eqnarray}
where the weighting function, $w$, may be specified or estimated as
described earlier in this section.  Then $z_i$ can then be computed
numerically as
 \begin{eqnarray}
 \label{eq:zi} z_i &=& \int_{0}^{\infty} \Psi_2^{\#}(y)\gamma_i(y)dy.
\end{eqnarray}
This form for the $z_i$ is simpler to evaluate than the double integral
form for these terms given earlier.
In a similar fashion we can write $q_{ij}$  as
\begin{eqnarray} \label{eq:qij} q_{ij} &=& \int_{0}^{\infty}
\Gamma_i(t) \Gamma_j(t) dt \nonumber\\ &=&
\int_{0}^{\infty}\int_{y}^{\infty}\int_{y}^{\infty} (s-y) \gamma_i(s)
(t-y) \gamma_j(t) ds dt dy \nonumber \\ &=&
\int_0^{\infty}\int_0^{\infty}\left[ s t \min[s,t] - {(s+t)
   \min[s,t]^2\over 2} + {\min[s, t]^3\over 3}\right]
\gamma_i(s) \gamma_j(t) dt ds \nonumber \\
\end{eqnarray}
 These integrals (\ref{eq:zi}) and (\ref{eq:qij}) can be computed
 explicitly and Matlab codes are available at website
 {\it http://www.stat.lsa.umich.edu/\~{}wangxiao} to do so.

 If we define $\hat{\bf \beta}$ as the solution to the quadratic
 minimization problem (see Equation (\ref{eq:kappa})), our estimates of
 $\Psi(y)$, $\Psi'(y)$, $\mu(r)$, and $M(r)$ can be denoted using
 analogous notation and are given by
$$\hat{\Psi}(y) = \sum^{m-1}_{k=1} \hat{\beta}_k \Gamma_k(y),$$
$$\hat{\Psi}'(y) = \sum^{m-1}_{k=1} \hat{\beta}_k \Gamma_k'(y),$$
$$\hat{\mu}(r) = -{1\over \pi^2}{{\hat{\Psi'}(r^2)}\over{{f}(r)}},$$
and,
$$\hat{M}(r) = \sum^{m-1}_{k=1} \hat{\beta}_k \left\{\left[r -
      r_{k-1}\right]^2_+ -\left[r - r_{m-1}\right]_+^2\right\}.$$
The latter two functions specify the projected radial velocity and mass
profiles, respectively.

 \begin{figure}[h]
 \begin{center}
   \resizebox{3in}{3in}{\includegraphics{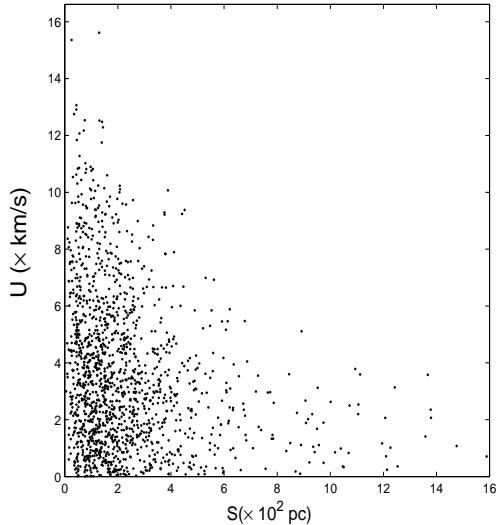}}
 \end{center}\caption{\label{fig:SU} \small A plot of the $(S,U)$ pairs
 used in one Monte Carlo realization described in Section 5.1.}
 \end{figure}

\begin{figure}[h]
 \begin{center}
   \resizebox{5in}{3in}{\includegraphics{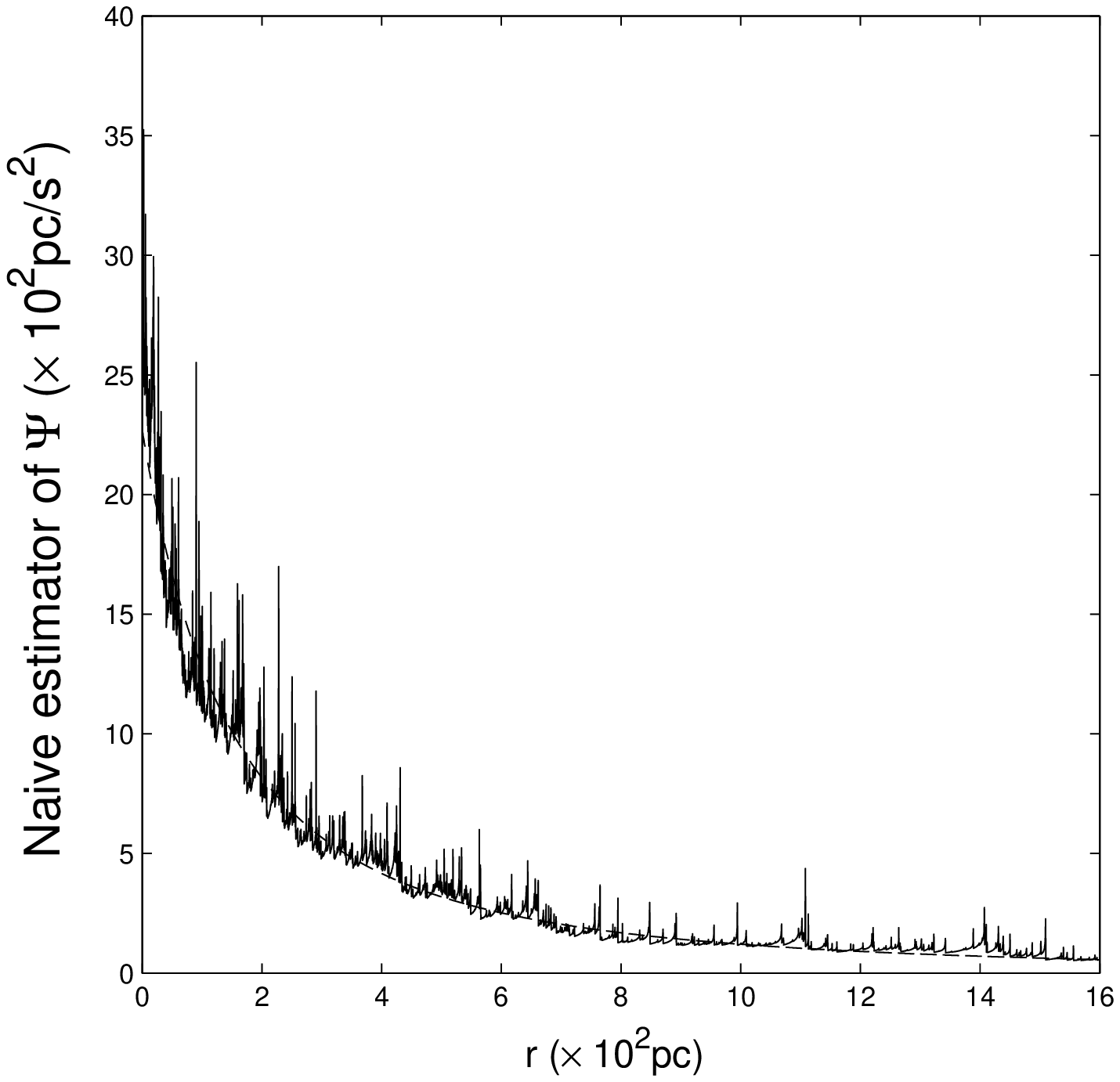}\hspace{0.2in}\includegraphics{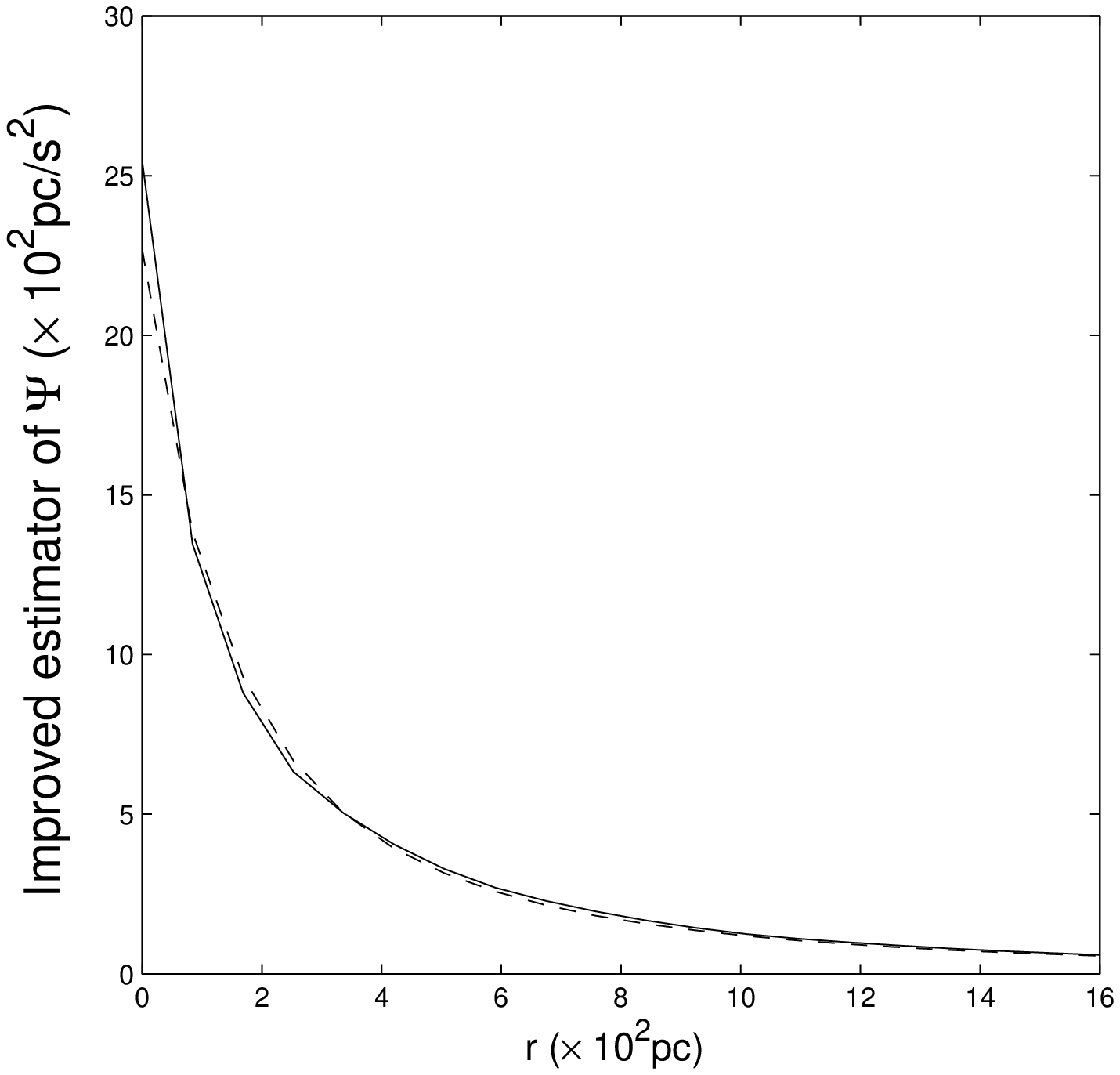}}
 \end{center}\caption{\label{fig:Psi}\small (Left)  The `naive' estimator
   $\Psi^\#$ as defined by Equation (\ref{eq:Psisharp}) and derived for
the Monte-Carlo simulation described in Section 5.1 and for the $(S,U)$ pairs shown in Figure \ref{fig:SU}.
(Right) The
   improved estimator $\hat{\Psi}$ derived from imposing the shape
   restrictions on the form of $M(r)$ (see Equations (\ref{eq:spln1})
   and (\ref{eq:spln2})) for the same simulated data set.  In both figures the solid line is the
   estimator and the dashed line is the true distribution calculated
   from the underlying Plummer model.}
 \end{figure}

 \begin{figure}[h]
 \begin{center}
   \resizebox{3in}{3in}{\includegraphics{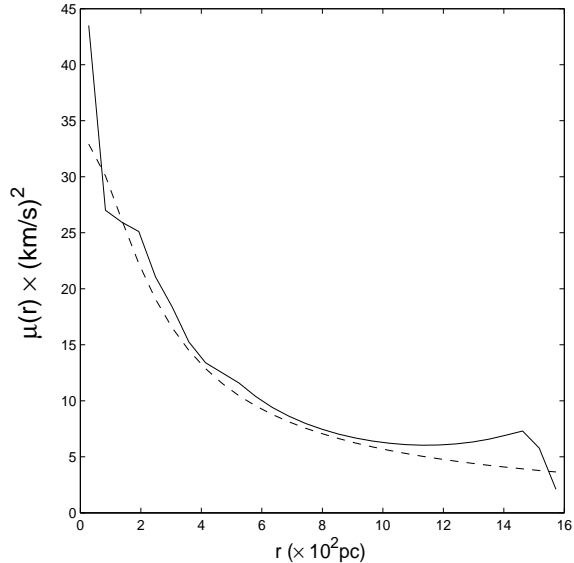}}
 \end{center}\caption{\label{fig:mu} \small The radial velocity profile,
   $\mu(r)$, derived from the Monte-Carlo realization based on the
 Plummer Model described in Section 5.1.  The actual profile
 calculated from the Plummer Model is shown as a dashed line.  Our
 estimator, $\hat{\mu}(r)$ is shown as a solid line.}
 \end{figure}

 \begin{figure}[h]
 \begin{center}
   \resizebox{3in}{3in}{\includegraphics{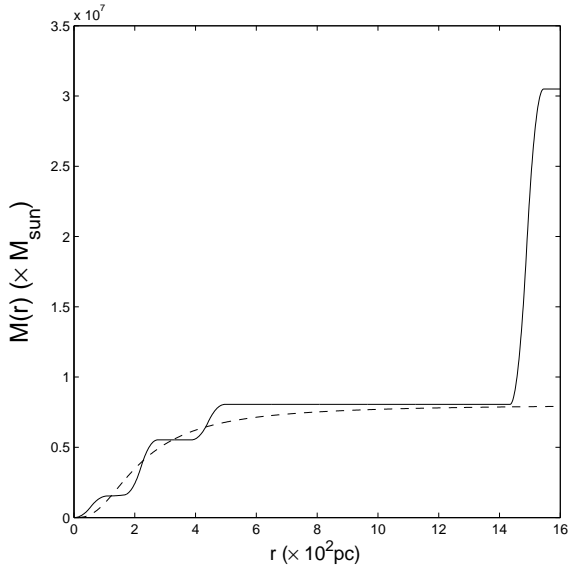}}
 \end{center}\caption{\label{fig:mass} \small The estimator
   of the mass distribution, $\hat{M}(r)$ for the Monte-Carlo
   simulation based on a Plummer model described in Section 5.1.  The
   piecewise nature of the estimator (see Equation (\ref{eq:spln1}) is
   apparent.  The dashed line is the true mass distribution for the
   underlying Plummer model.}
 \end{figure}

\begin{figure}[h]
 \begin{center}
   \resizebox{3in}{3in}{\includegraphics{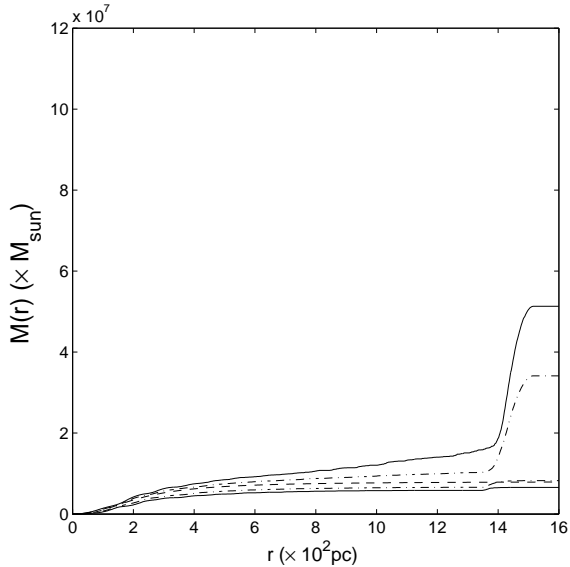}}
 \end{center}\caption{\label{fig:masscb} \small A plot of the
   95\%\ (solid lines) and 68\%\ (inner dot-dashed lines)
   confidence bounds on the mass distribution, $M(r)$ (see Figure
   \ref{fig:mass}).  The actual mass distribution from the Plummer
   model is shown as a dashed line.  The strong positive bias at large
   $r$ (especially for $M(r)$) is evident for $r > 14$ where the
   number of data points constraining the mass distribution is very
   low (about 1\%\ of the entire sample lies outside $r \sim 14$).}
 \end{figure}

 \begin{figure}[h]
 \begin{center}
   \resizebox{3in}{3in}{\includegraphics{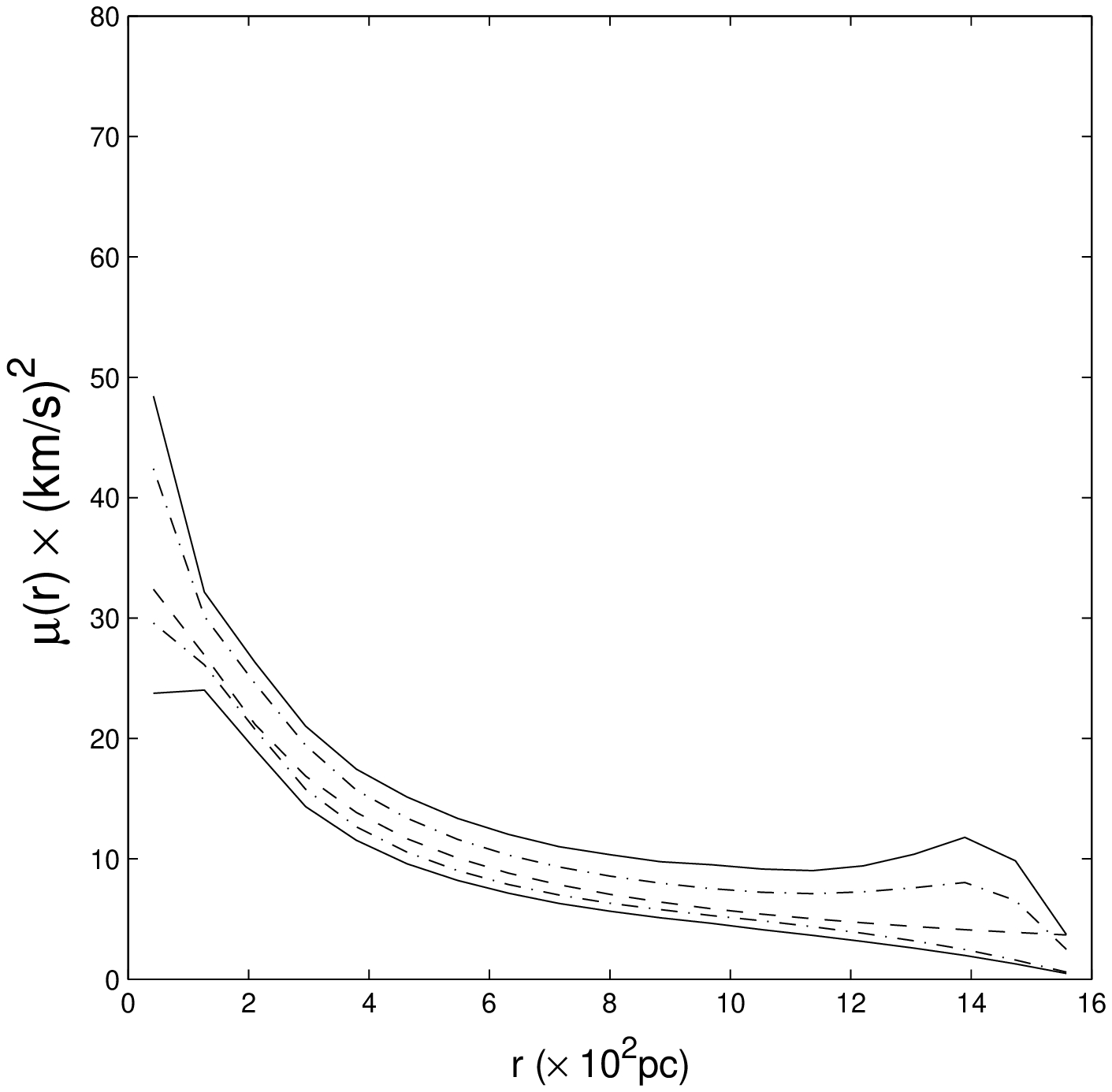}}
 \end{center}\caption{\label{fig:mucb} \small The
   95\%\ (solid lines) and 68\%\ (dot-dashed lines)
confidence bounds on the velocity dispersion profile
   estimator, $\hat{\mu}(r)$, along with the actual dispersion profile
   for the Plummer model (dashed line).}
 \end{figure}

 \section{The Mass Profile: Simulations and a Trial Application}

 \subsection{Monte-Carlo Simulations}

 To illustrate an application of our approach, we have drawn a sample
 of 1500 $(S,U)$ pairs from a Plummer model with $b = 200\times 3.1\times 10^{15}\ km^3/sec^2$ and
 $\sigma_i^2 = 1$ (see Figure \ref{fig:SU}).  The naive estimator
 $\Psi^\#$ may be computed from these data using Equation
 (\ref{eq:Psisharp}) and it is shown in Figure \ref{fig:Psi}, along
 with the improved estimator $\hat{\Psi}$.  The need for the
improved estimator for $\Psi$ is clear since, as expected,
 $\Psi^\#$ is a highly irregular function which reacts strongly to the
 presence of individual data points.  The estimated velocity
 dispersion profile, $\hat{\mu}(r)$ is presented in Figure
 \ref{fig:mu} and the estimated mass profile, $\hat{M}(r)$, is
 illustrated in Figure \ref{fig:mass}.  In all cases, the dotted line
 represents the true functions computed directly from the Plummer
 model from which the $(S,U)$ pairs were drawn.  In these figures $n = 1500$, $m = 30$, and $w \equiv 1$.

\begin{figure}[p]
\begin{center}
 \begin{minipage}[c]{0.75\textwidth}
 \begin{center}
 \resizebox{5in}{1.5in}
{\includegraphics{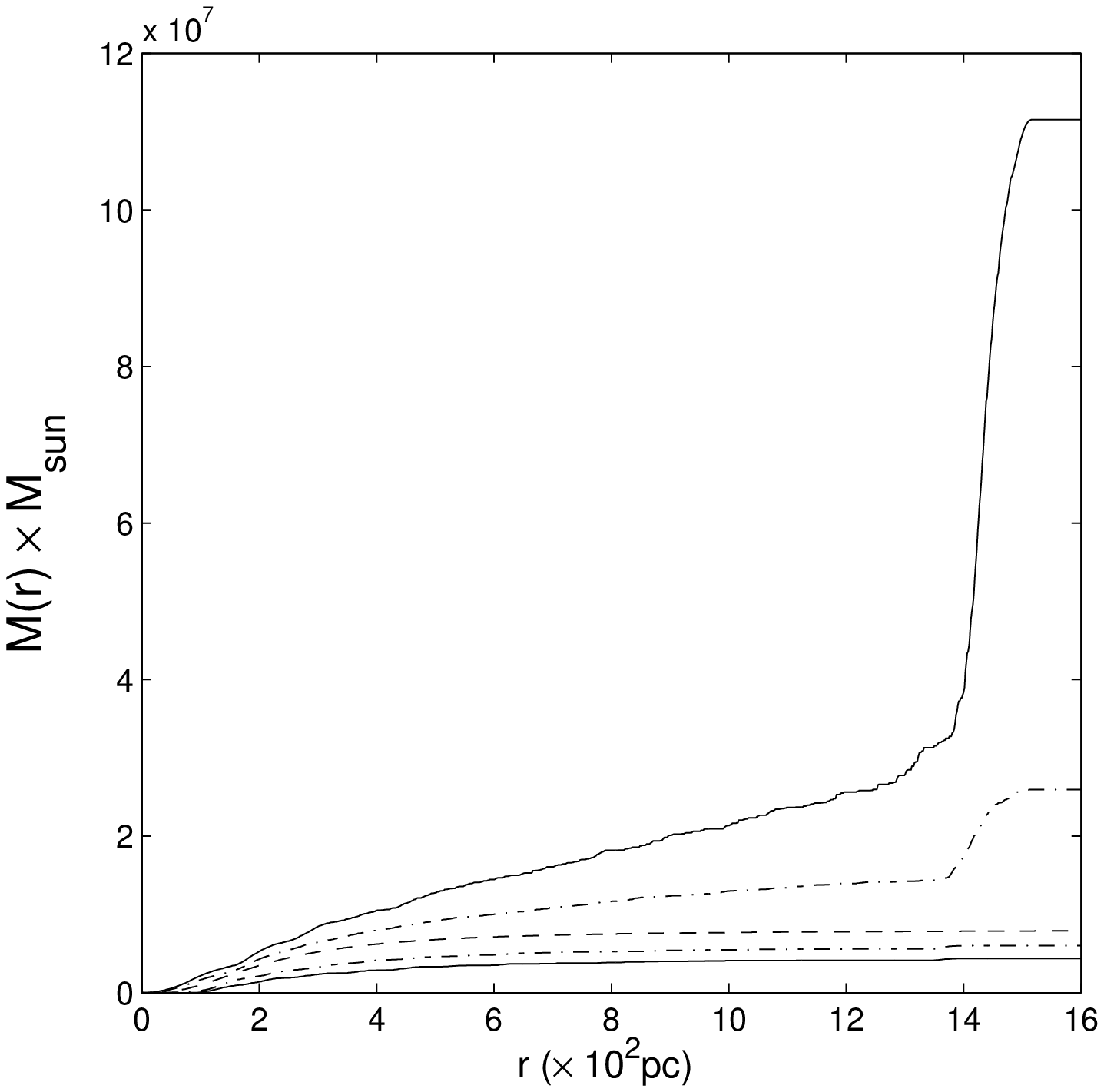}\includegraphics{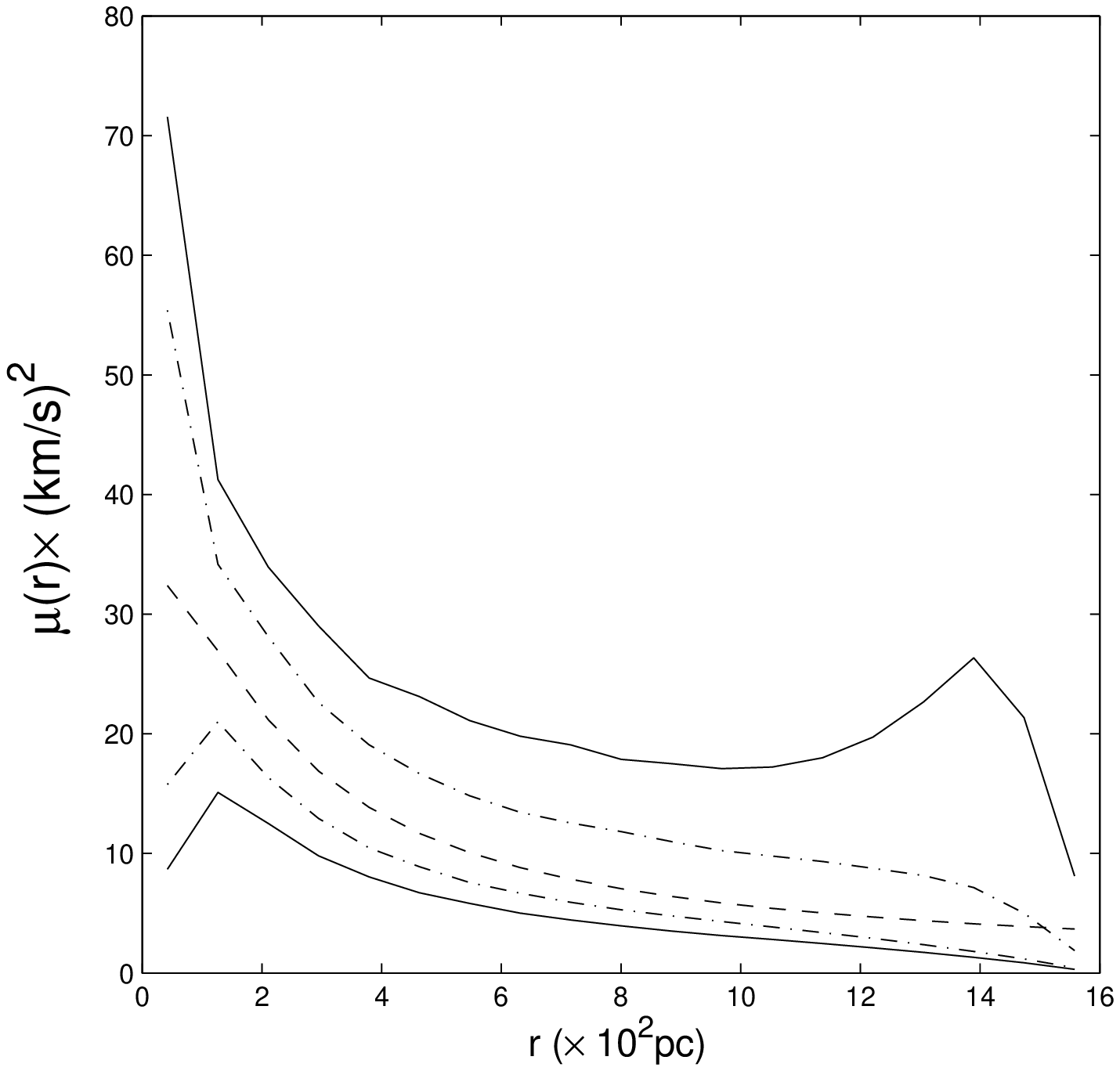} }
 \end{center}
\end{minipage}
 \begin{minipage}[c]{0.75\textwidth}
 \begin{center}
 \resizebox{5in}{1.5in}
{\includegraphics{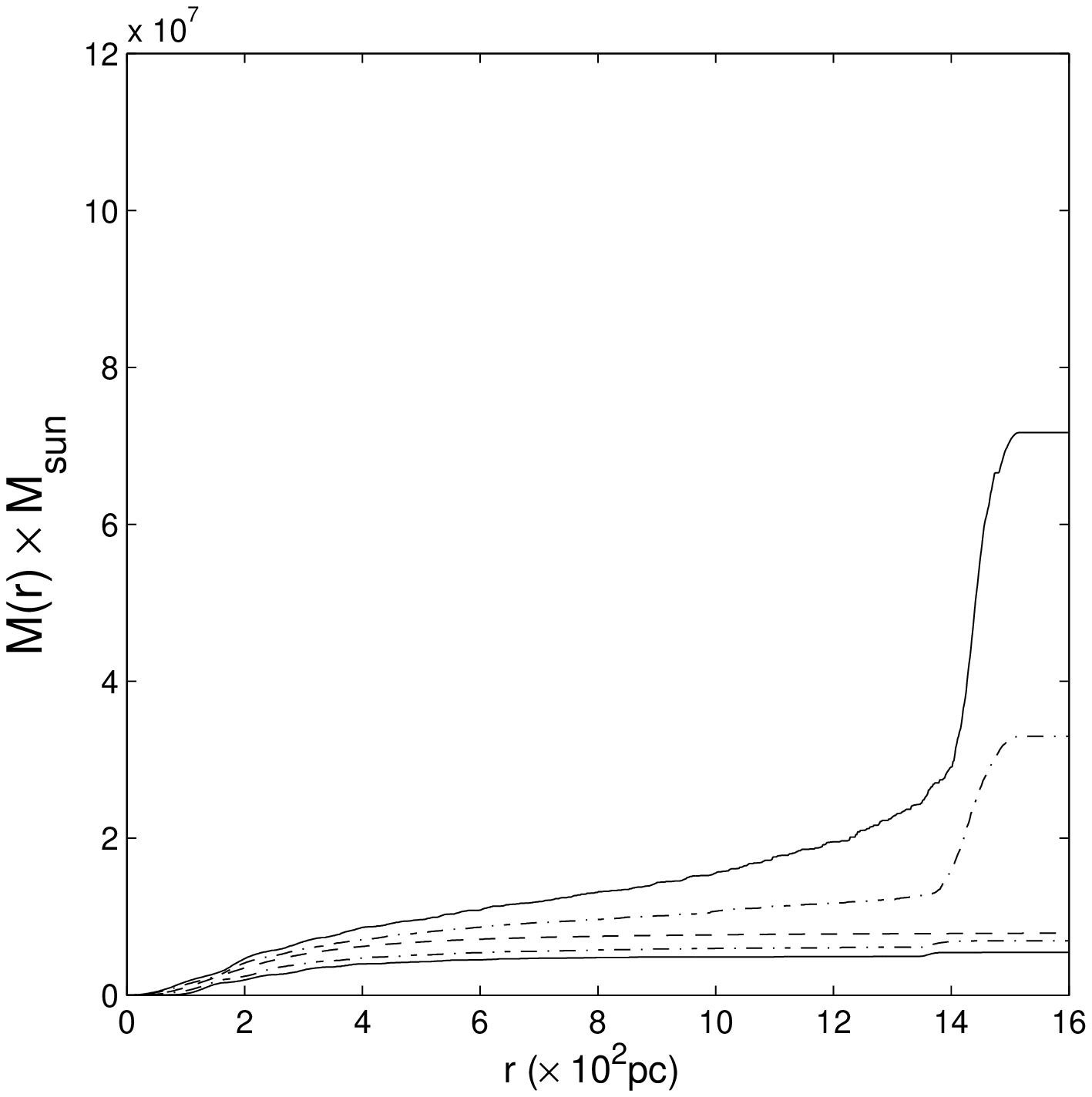}\includegraphics{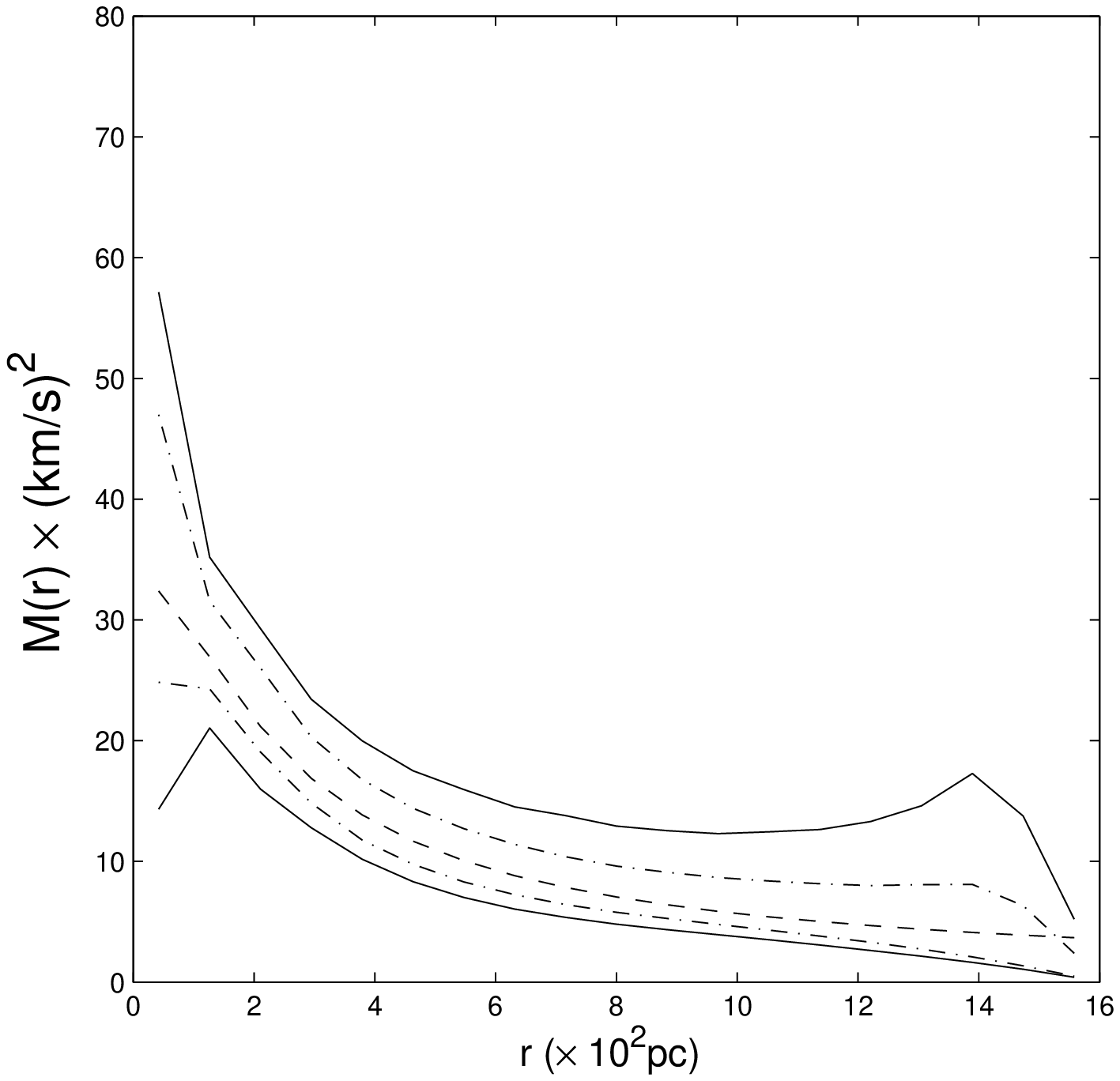} }
 \end{center}
\end{minipage}
 \begin{minipage}[c]{0.75\textwidth}
 \begin{center}
 \resizebox{5in}{1.5in}
 {\includegraphics{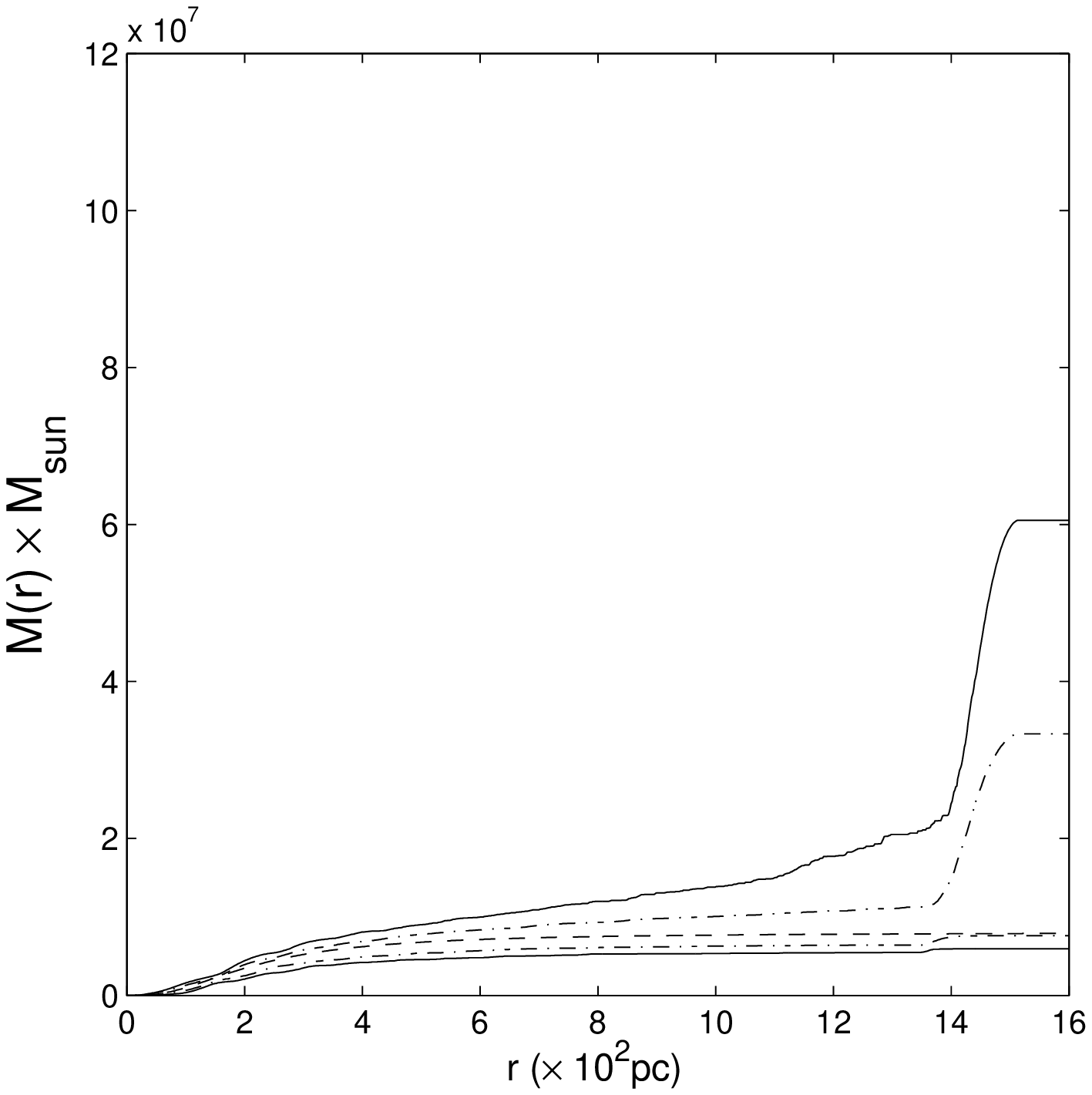}\includegraphics{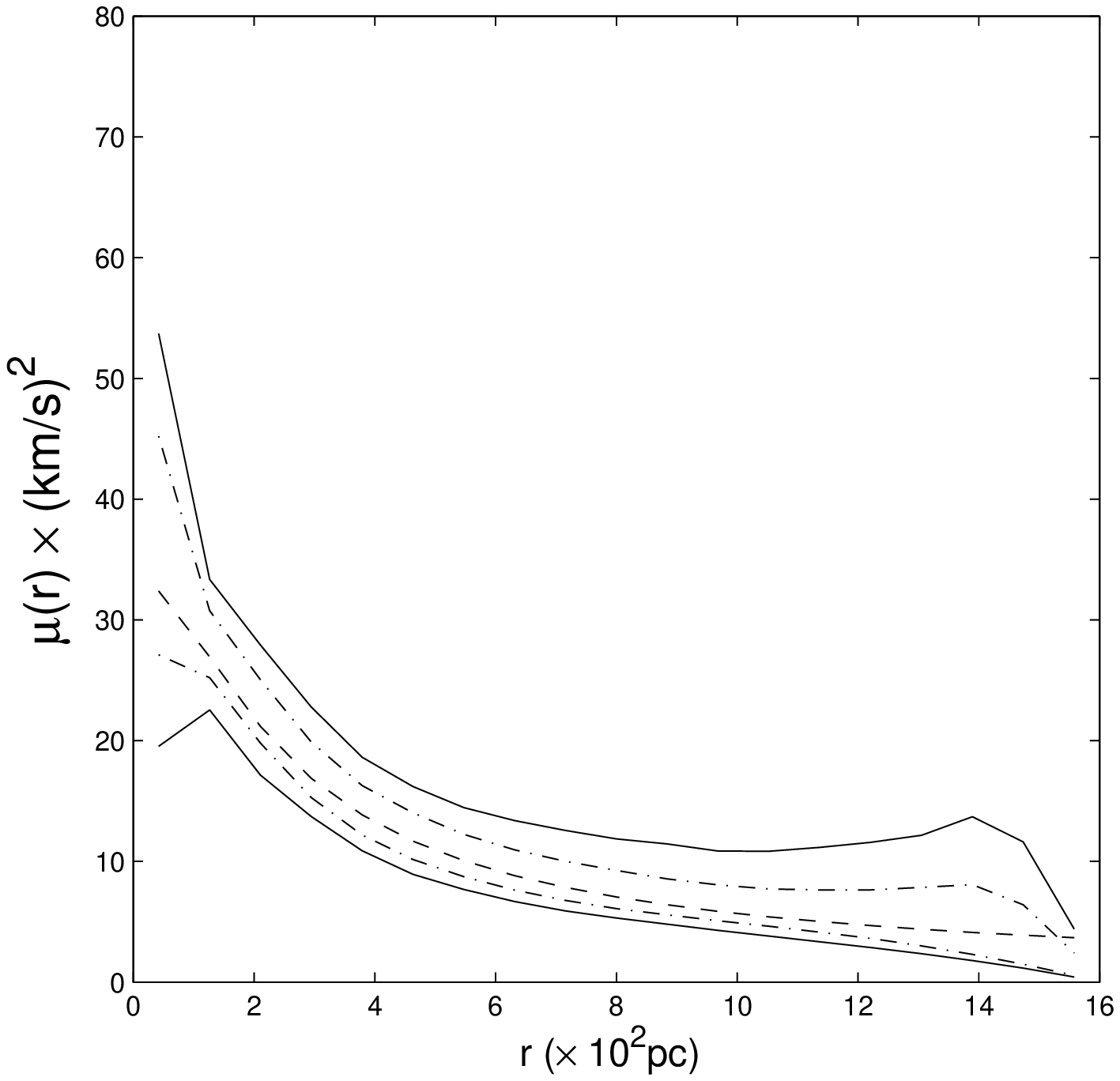} }
 \end{center}
\end{minipage}
 \begin{minipage}[c]{0.75\textwidth}
 \begin{center}
 \resizebox{5in}{1.5in}
{\includegraphics{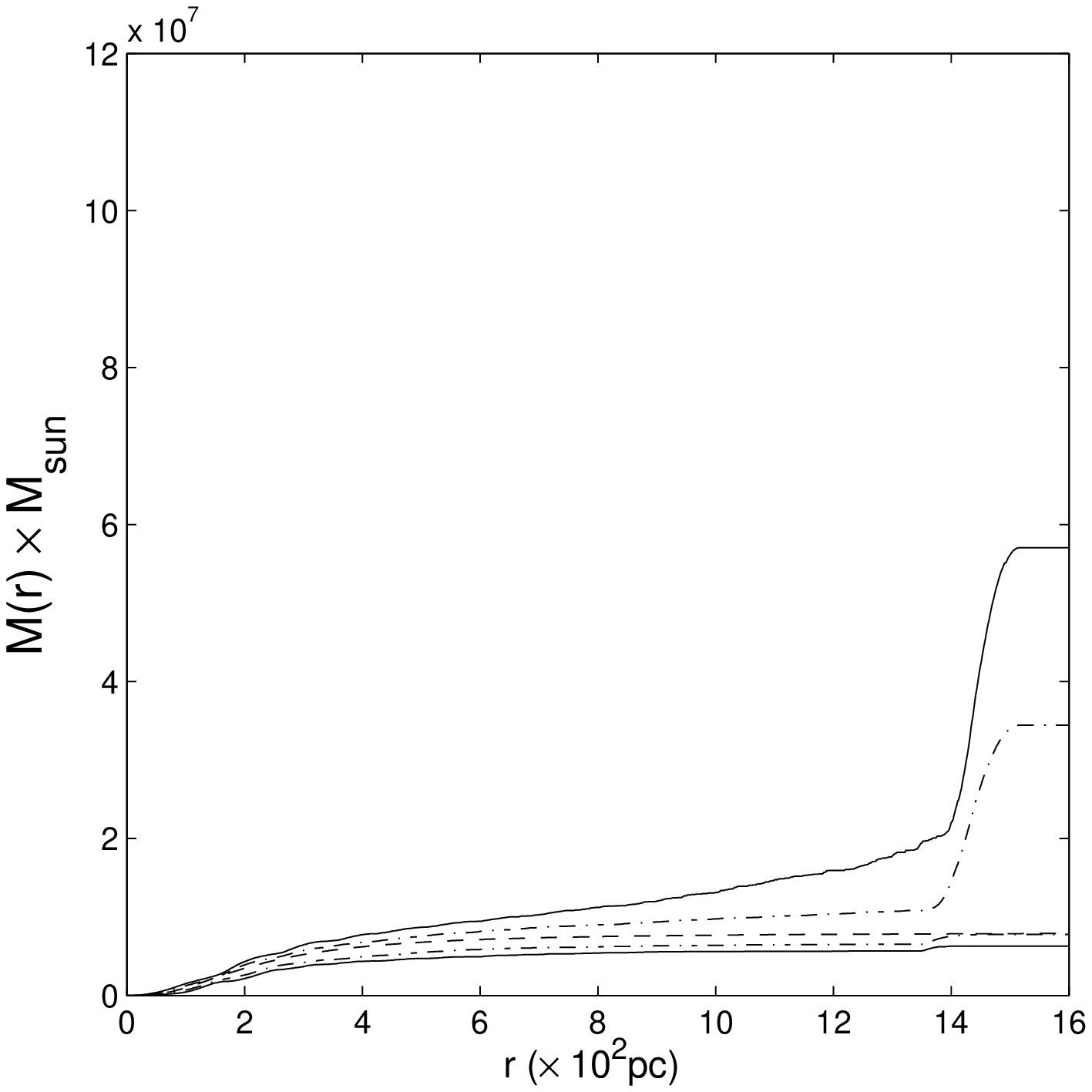}\includegraphics{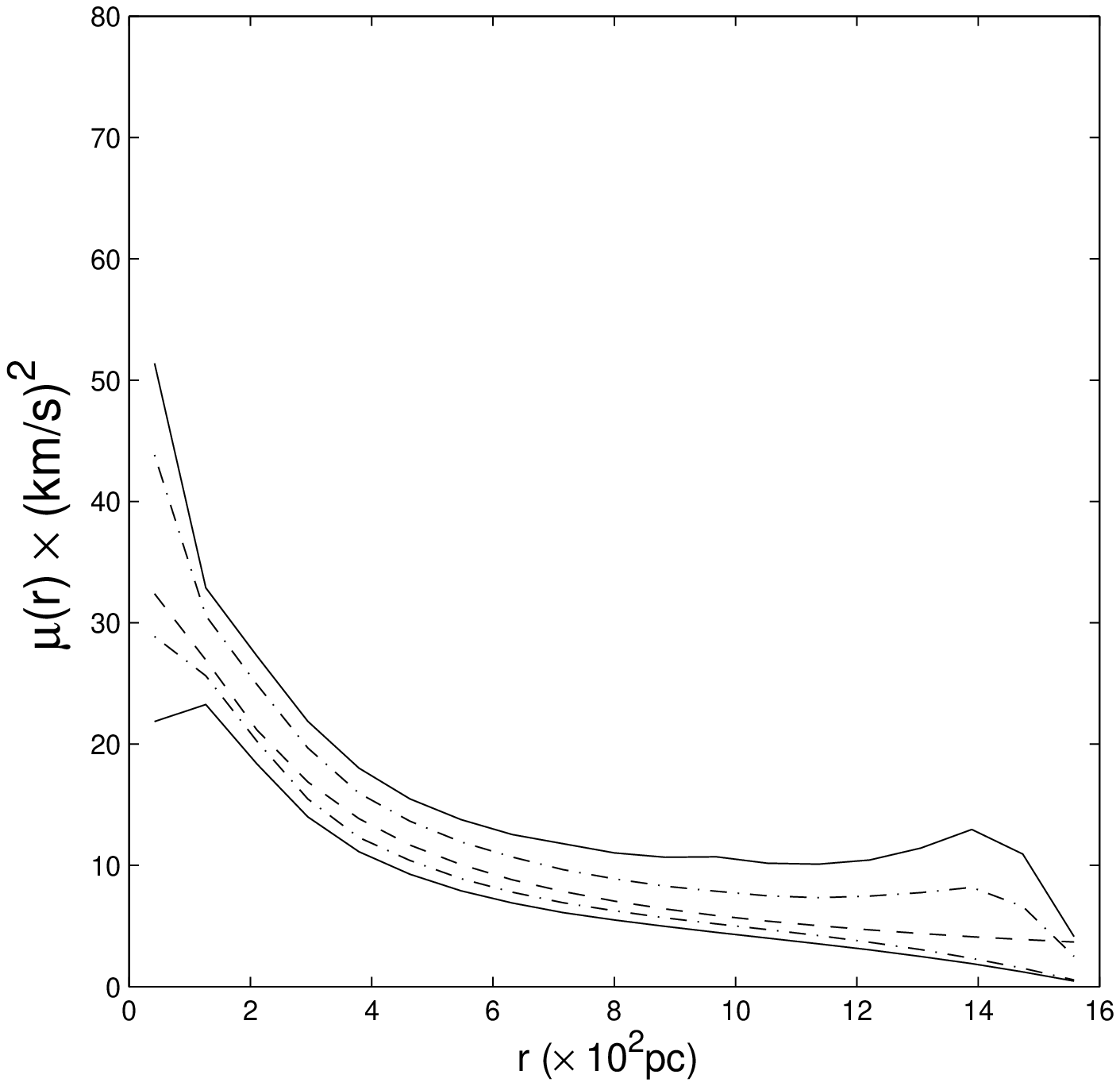} }
 \end{center}
\end{minipage}
\end{center}\caption{\label{fig:diffsample} \small The $95\%$ (solid lines) and $68\%$ (dot-dashed lines) confidence bounds for different sample sizes of stars (from top, for 100, 400, 700 and 1000 stars)
  derived from a Plummer Model using Monte Carlo methods.  The
  lefthand figures show the confidence bounds for the mass
  distribution estimator, $\hat{M}(r)$, and the righthand panels show
  the confidence bounds for the velocity dispersion profile estimator,
  $\hat{\mu}(r)$.  In each panel, the dashed line is the true
  distribution for the Plummer model. }
\end{figure}

The last plataux on the right in Figure \ref{fig:mass} ($r > 14$) deserves comments, because it illustrates an
important feature of the problem in general and our method in particular: Estimates become unreliable for large
values of $r$, because the data become sparse, and shape restricted methods are especially prone to problems
such problems--e.g. Woodroofe and Sun (1993).   Deciding exactly where reliability ends is difficult, but the
confidence bands of Figures \ref{fig:masscb} and \ref{fig:diffsample} provide some indication.  In the example
of Figure 6, they correctly predict that reliability deteriorates abruptly for $r > 14$ and, interestingly, make
the same prediction for a range of sample sizes.

\subsection{Assessing the Estimation Error}

There are several sources of error in our derived velocity dispersion
and mass profile estimators which may be divided into the broad
classes of modeling error, systematic error, and statistical error.
While quite general, our assumptions also represent some
over-simplification and these too introduce uncertainty to our
results.  The assumption of spherical symmetry, for example, ignores
the known ellipticities seen in most dSph galaxies (IH95, Mateo 1998).
But even if we assume our working assumptions were all correct,
systematic error will develop from our approximations of distributions
as step functions, piecewise linear functions, and quadratic splines.
On general grounds, this error will be small provided that the
underlying true functions are smooth, so we feel justified in ignoring
this source of systematic error.  We have also ignored the statistical
error in the estimation of $f$ in Section \ref{sect:h}, because the
sample sizes in IH95 are so large.  That leaves the statistical error
in the kinematic data as the principle source of uncertainty in our
estimation of $\Psi$ and the estimators of $\mu$ and $M$.

To quantify the magnitude of this error,
let $e(r) = \hat{M}(r)-M(r)$ denote the error committed when
estimating $M(r)$, and let $z_{\alpha}(r)$ denote the $100\alpha^{th}$
percentile of its distribution distribution, so that $P[e(r) \le
z_{\alpha}(r)] = \alpha$.  Then
$$
P[\hat{M}(r)-z_{1-\alpha}(r) \le M(r) \le \hat{M}(r)-z_{\alpha}(r)] =
1-2\alpha,
$$
and $[\hat{M}(r)-z_{1-\alpha}(r),\hat{M}(r)-z_{\alpha}(r)]$ is a
$100(1-2\alpha)$ percent confidence interval for $M(r)$.  Within the
context of Plummer's Model, the $z_{\alpha}(r)$ can be obtained from a
Monte Carlo simulation.  If we generate $N_{MC}$ samples from
Plummer's Model, we obtain $N_{MC}$ values for the errors $e(r)$,
which we denote as $e_1(r),\cdots,e_N(r)$.  Then $z_{\alpha}(r)$ may
be estimated by the value of $z$ for which the $N\alpha$ of the
$e_i(r)$ are less than or equal to $z$.  We applied this procedure to
a grid of $r$ values with $N_{MC} = 2000$ and $\alpha = 0.025$ and
then connected the resulting bounds to in a continuous piecewise
linear fashion to obtain the 95\%\ and 68\%\ confidence bounds shown
in Figure \ref{fig:masscb}.  The 95\%\ and 68\%\ confidence bounds for
$\mu$ shown in Figure \ref{fig:mucb} were obtained in a similar
manner.

The procedure just outlined requires knowing the true $M(r)$ and the distributions of $Y$ and $V_3$ from the
Plummer model.  But the exercise is still useful since it shows the intrinsic accuracy of the estimation
procedure.  For example, when the sample size is $1500$, the estimator of $M(r)$ has very little accuracy for $r
\ge 14$. Somewhat surprisingly, the estimator for $M(r)$ retains reasonable accuracy for $r \le 14$ despite the
lack of much data beyond $r = 14$ (see Figure \ref{fig:counts}). We have also explored how our estimates fare
with significantly smaller samples of tracers with known positions and velocities (i.e., fewer $(S,U)$ pairs).
Figure \ref{fig:diffsample} shows $95$ percent confidence bounds for the estimates of $M(r)$ and $\mu(r)$ based
on sample sizes $100$, $400$, $700$ and $1000$ stars.  As one might expect, the confidence bounds expand as the
sample size decreases, but, interestingly, the radial location of where the estimator performs poorly only very
slowly migrates to smaller values of $r$ as the sample shrinks.

The requirement that we know $M(r)$ and the distribution of $Y$ and
$V_3$ ahead of time in order to estimate the confidence bounds can be
avoided by using a bootstrap procedure.  In its simplest form, the
bootstrap can be implemented as follows: Given a sample of $(Y,V_3)$
values (corresponding to the observable projected position and radial
velocity for each star in a kinematic sample), first compute
$\hat{M}(r)$.  Then generate $N_b$ samples with
replacement\footnote{Drawing from a sample `with replacement' means
  that for an original sample of $m$ objects, one selects a new sample
  of $m$ values from the original values $m_i$.  The new sample
  differs from the original in that each value $m_i$ is returned to
  the sample before choosing the next value $m_{i+1}$.  In this way,
  some values from the original sample will invariably be chosen more
  than once to form the bootstrap sample.}  from the given sample and
compute estimators $\hat{M}_1^*(r),\cdots, \hat{M}_N^*(r)$ for each of
these samples.  Let $\hat{e}_i^*(r) = \hat{M}_i^*(r)-\hat{M}(r)$ and
estimate the percentiles of $\hat{e}_i^*(r)$ as described above.
Letting $\hat{z}_{\alpha}^*(r)$ denote the estimated $100\alpha^{th}$
percentile, $[\hat{M}(r)-\hat{z}_{1-\alpha}^*(r),
\hat{M}(r)-\hat{z}_{\alpha}^*(r)]$ is the bootstrap confidence
interval for $M(r)$.  For more details, see Efrom and Tibshirani
(1993) who provide a good introduction to bootstrap methods.  This
approach offers a practical way of assessing the statistical errors in
$\hat{\mu}(r)$ and $\hat{M}(r)$ of an actual kinematic sample.

There is also the possibility of obtaining large-sample approximations
to the error distribution.  From (\ref{eq:zi}), it is clear that the
$z_i$ are approximately normally distributed in large samples.  This
suggests that the distribution of $\hat\beta$ should be approximately
the same as the distribution of a shape-restricted estimator in a
normal model.  Unfortunately, calculating the latter is complicated.
Some qualitative features are notable, however.  Monotone regression
estimators are generally quite reliable away from the end points, but
subject to the {\it spiking problem} near the end points.  For
non-decreasing regression, the estimator has a positive bias near
large values of the independent variable (in this case, $r$), and a
negative bias at small values of the independent variable (i.e.  for
$r \sim 0$).  The positive bias at large $r$ is very clearly seen in
Figure \ref{fig:masscb} and is exacerbated by the sparsity of data.
The negative bias near $r = 0$ is not evident because it is
ameliorated by two effects. First, the mass distribution is known to
be always positive ($M(r) \ge 0$) and goes to zero at $r = 0$ ($M(r=0)
= 0$).  This eliminates bias at $r = 0$.  Second, our assumption
(\ref{eq:spln3}) requires $\hat{M}(r)$ to be of order $r^2$ for small
$r$, while $M(r)$ is -- on physical grounds -- known to approach zero
as $r^3$ for small $r$.

\begin{figure}[h]
\begin{center}
\begin{minipage}[c]{0.80\textwidth}
 \begin{center}
\resizebox{3in}{3in} {\includegraphics{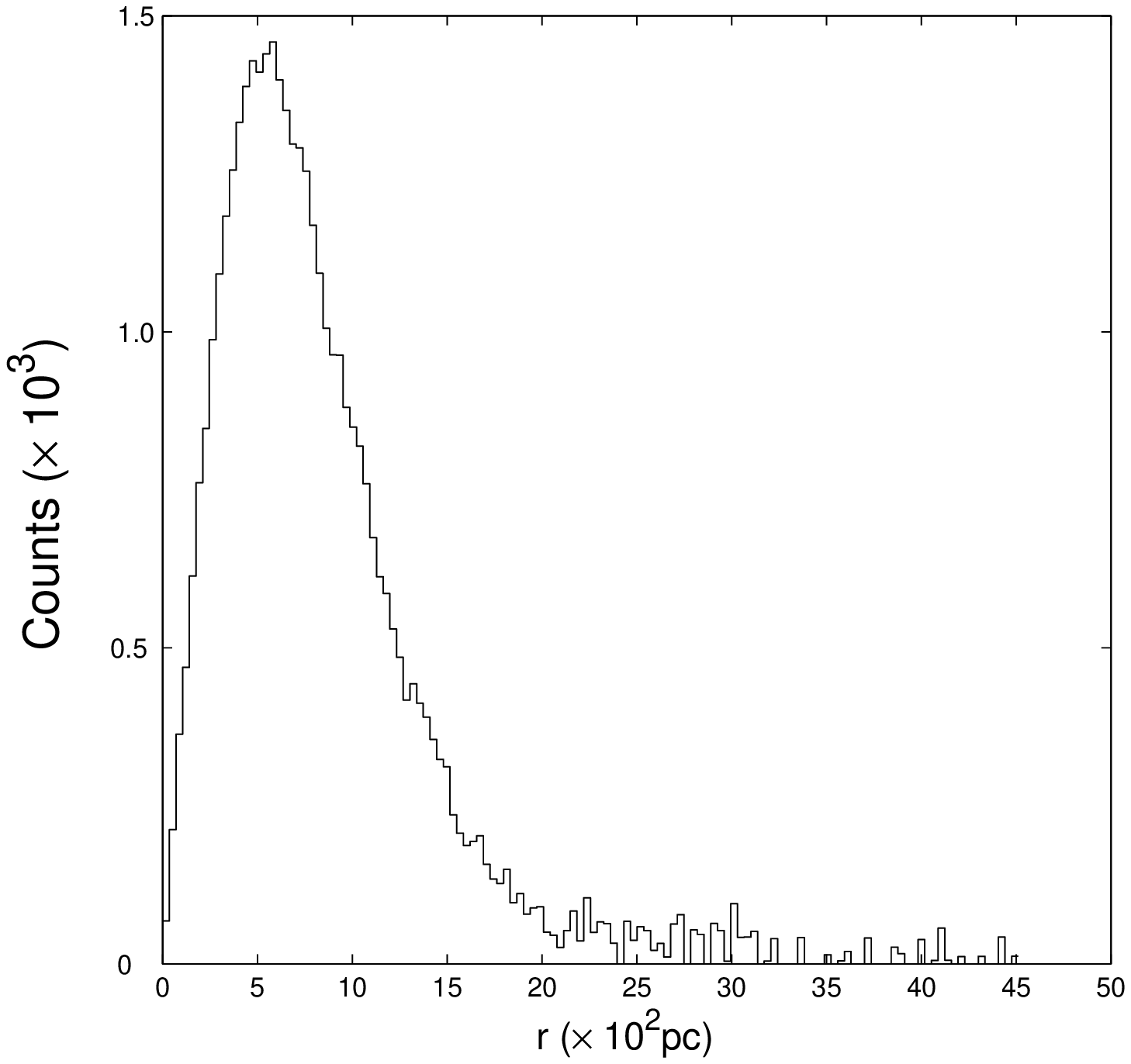} }
 \end{center}
 \end{minipage}
 \begin{minipage}[c]{0.80\textwidth}
 \begin{center}
\resizebox{5.5in}{3in} {\includegraphics{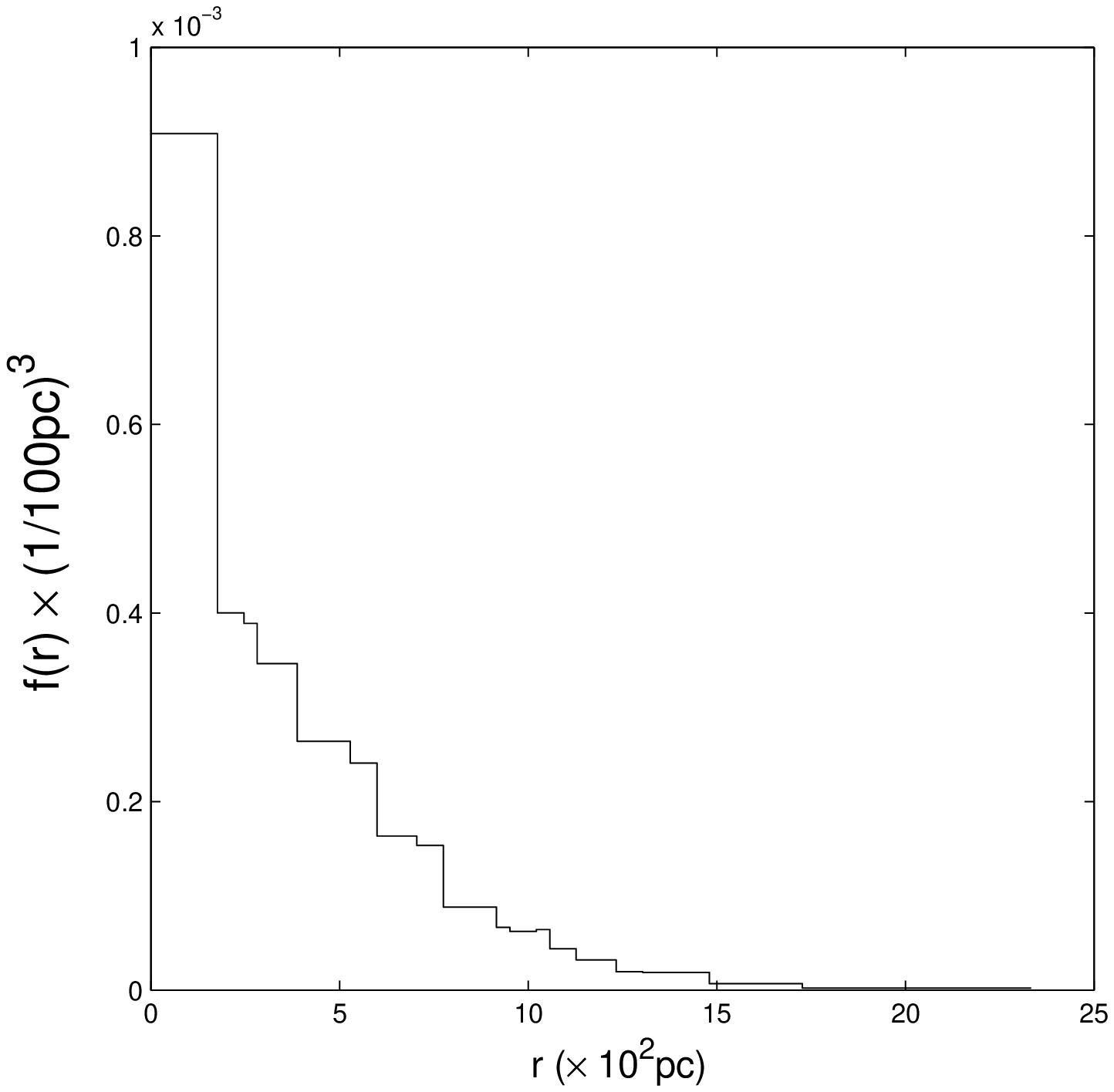} \hspace{.2in}\includegraphics{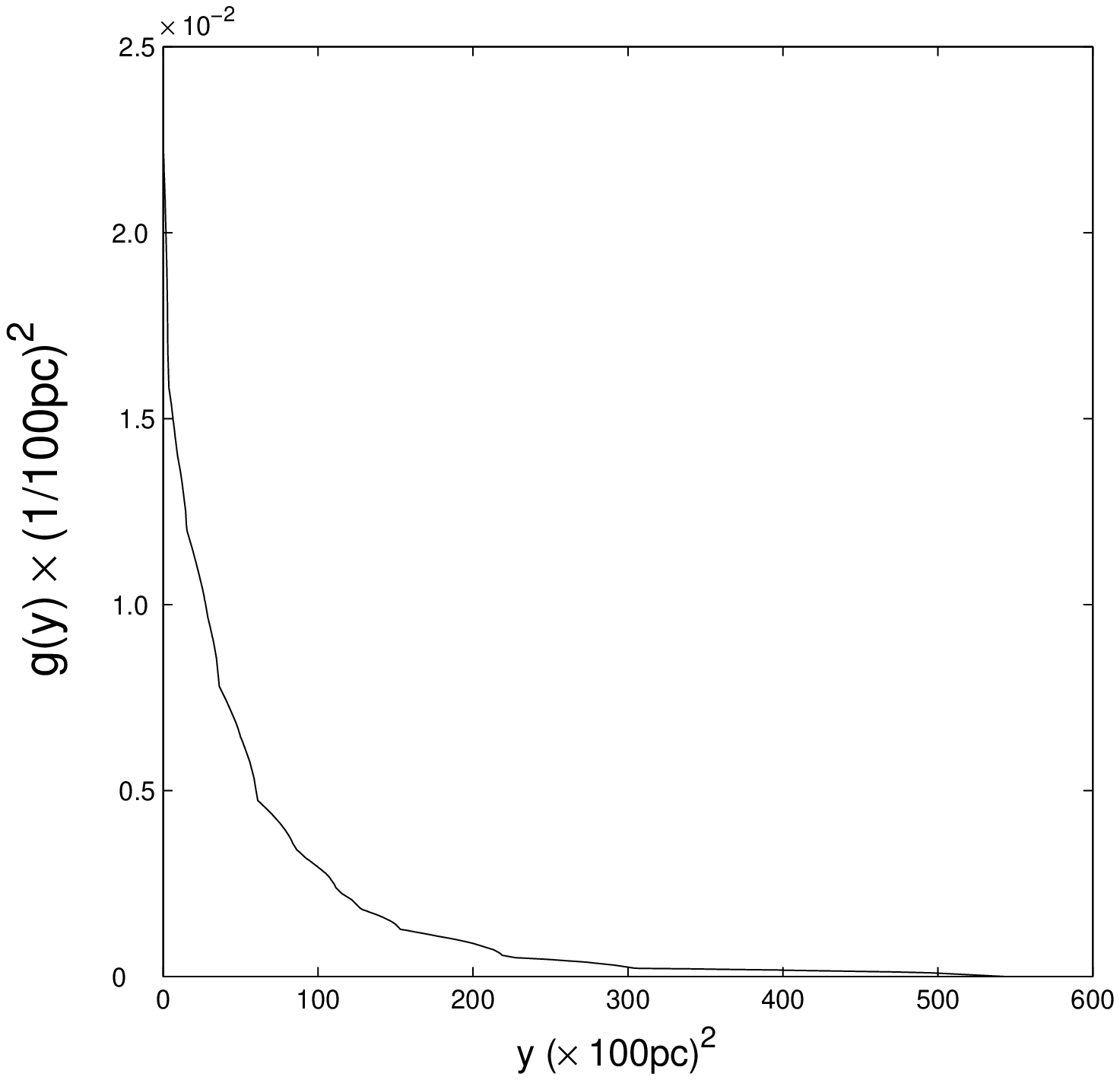}}
 \end{center}
 \end{minipage}
 %\begin{minipage}[c]{0.75\textwidth}
 %\begin{center}
%\resizebox{4.5in}{2.7in} {\includegraphics{gy-paper-fornax.eps} }
% \end{center}
% \end{minipage}
 \end{center}\caption{\label{fig:fornax1} \small Observed and derived
   distributions from projected counts data in (Irwin and
   Hatzidimitriou, 1995). Top: Histogram of counts from IH95. Bottom: The inferred three-dimensional distribution of stars in Fornax
   assuming spherical symmetry, $\hat{f}(r)$; The projected
   radial distribution of stars in Fornax, $\hat{g}_{\bf Y}(y)$.}
\end{figure}

\begin{figure}[h]
\begin{center}
\begin{minipage}[c]{0.80\textwidth}
\begin{center}
\resizebox{3in}{3in} {\includegraphics{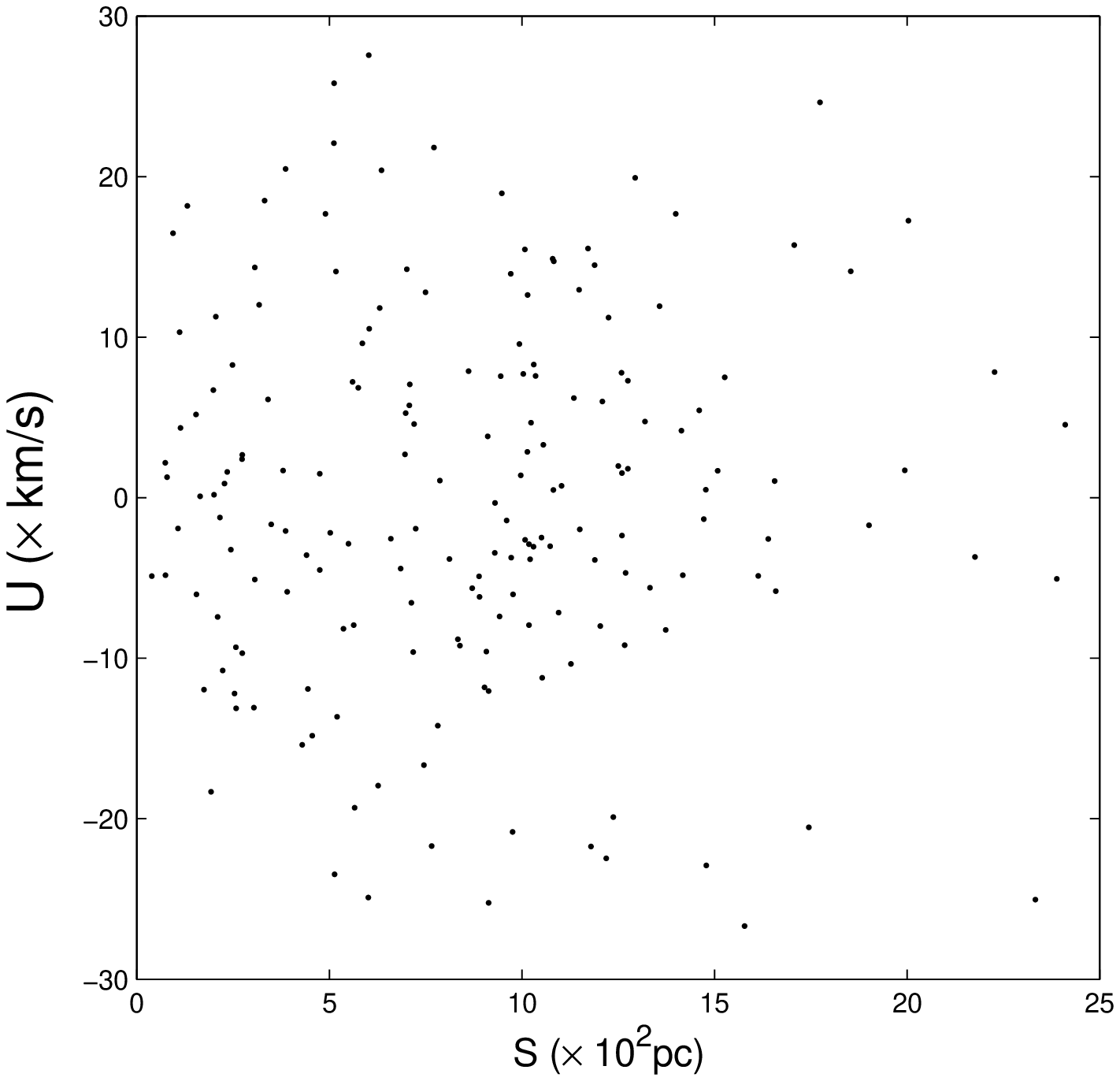} }
\end{center}
\end{minipage}
\begin{minipage}[c]{0.80\textwidth}
\begin{center}
\resizebox{5.5in}{3in} {\includegraphics{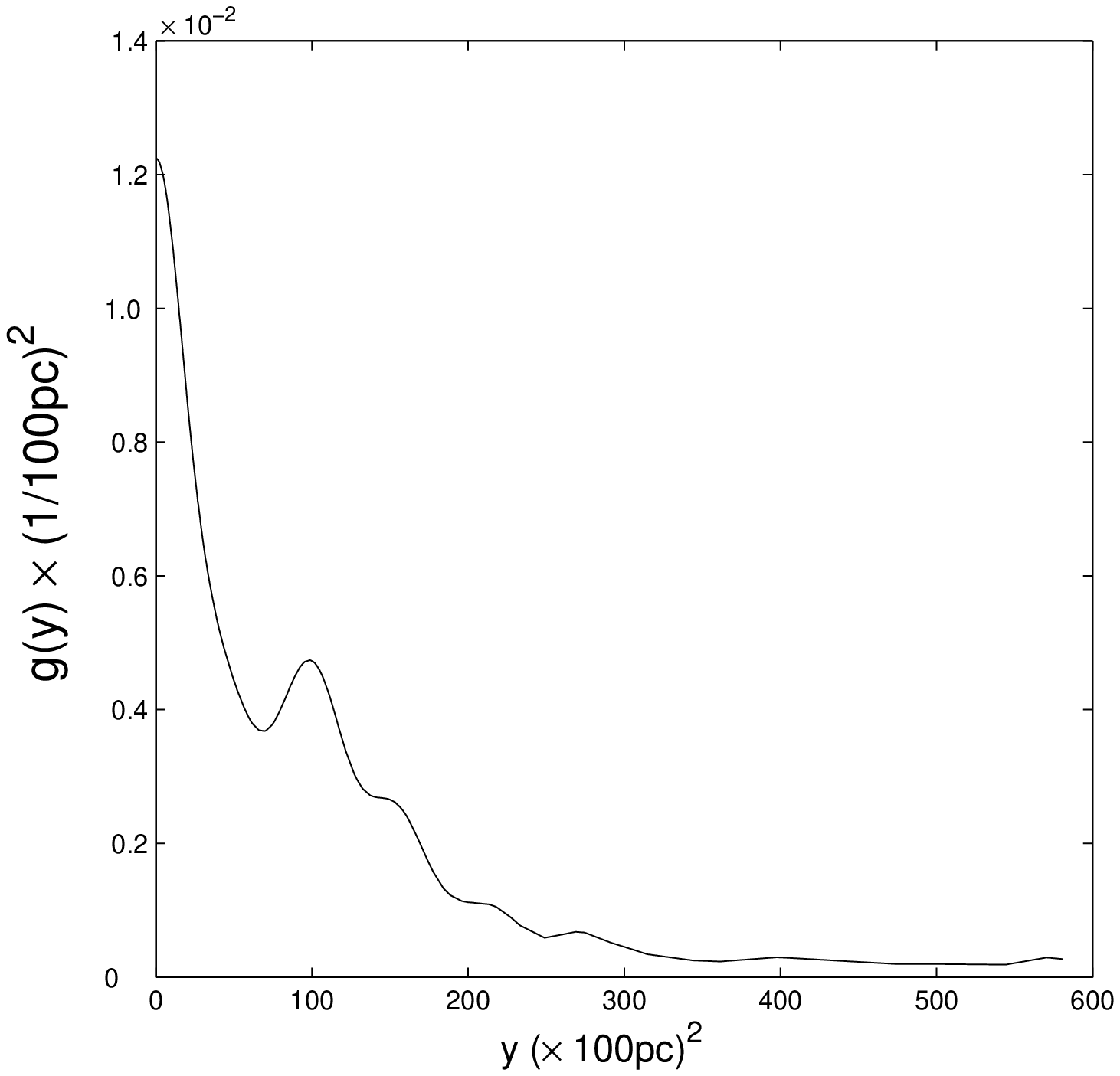}\hspace{.2in}\includegraphics{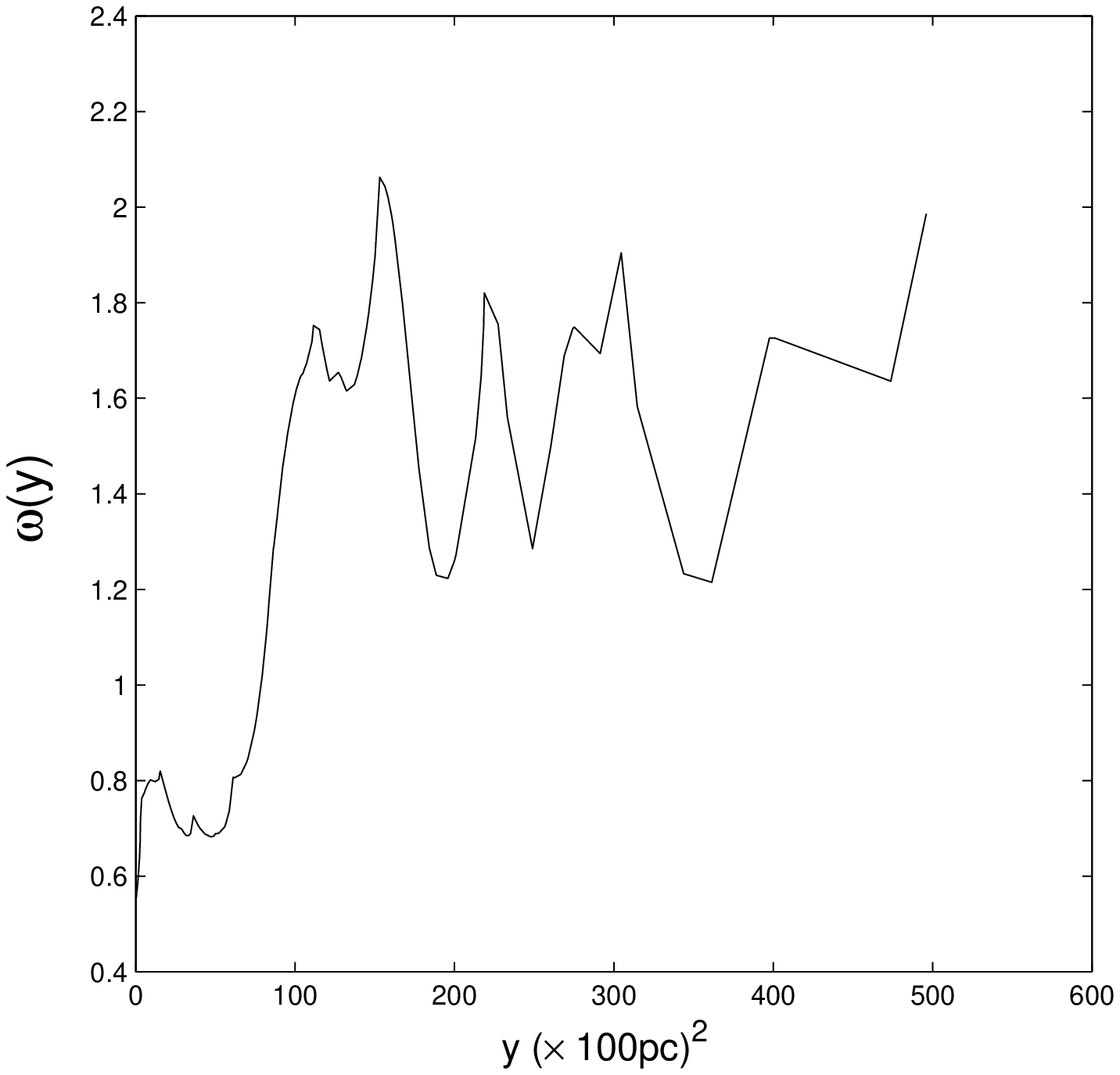}}
\end{center}
\end{minipage}
%\begin{minipage}[c]{0.75\textwidth}
%\begin{center}
%\resizebox{4.5in}{2.7in} {\includegraphics{selection-fornax.eps} }
%\end{center}
%\end{minipage}
\end{center}\caption{\label{fig:fornax21} \small Observed and derived
  distributions from the kinematic dataset for the Fornax dSph galaxy
  (Walker et al. 2004).  Top: The distribution of
  radial velocity/projected radial positions for the stars in the
  sample; Bottom: The projected radial distribution of stars in Fornax,
  $\hat{g}_{\bf Y}^*(y)$;  The estimated selection function
  $\hat{\omega}$;}
\end{figure}

\begin{figure}[h]
\begin{center}
%\begin{minipage}[c]{0.75\textwidth}
\begin{center}
\resizebox{5.5in}{3in} {\includegraphics{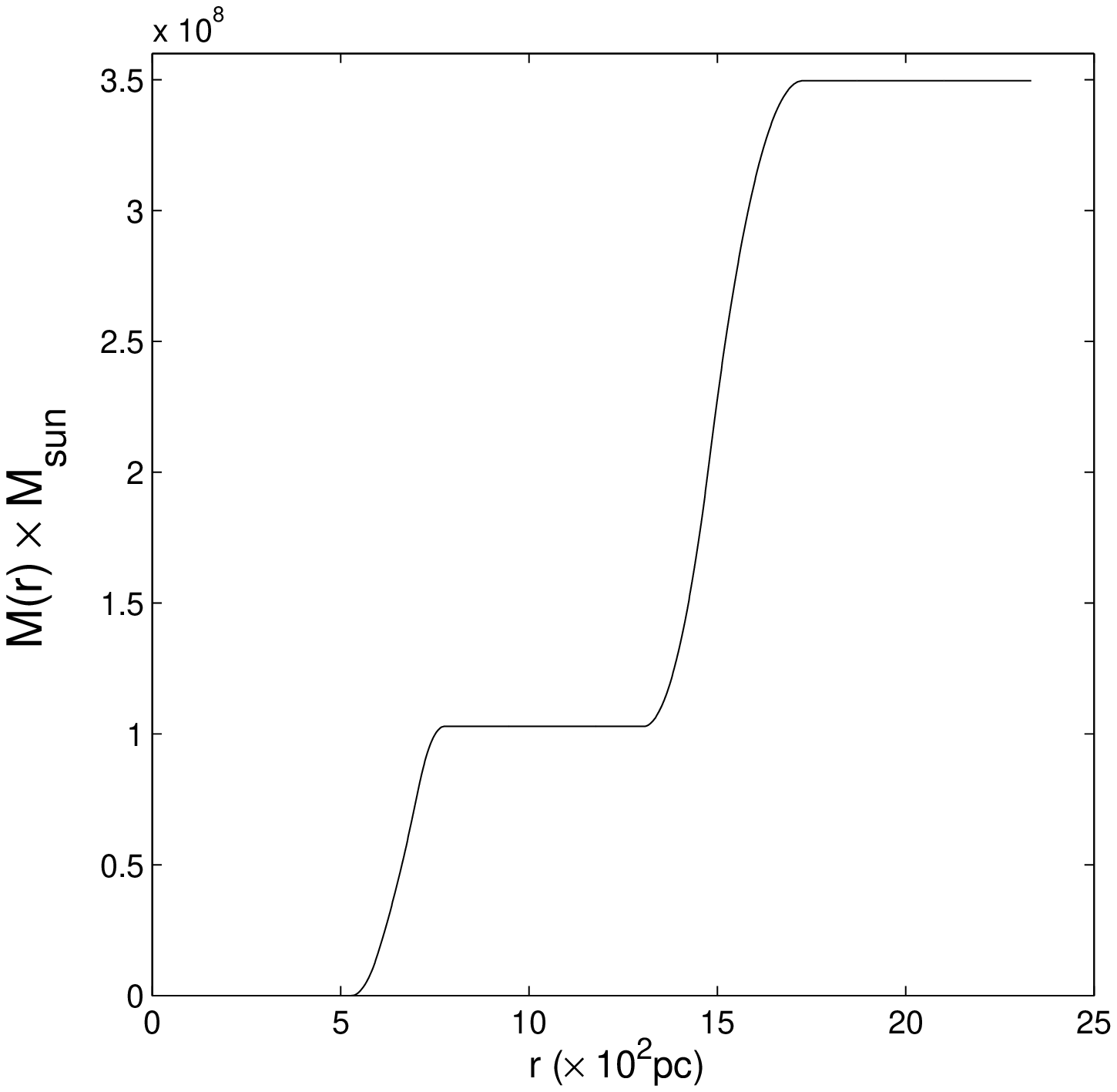} \hspace{.2in}\includegraphics{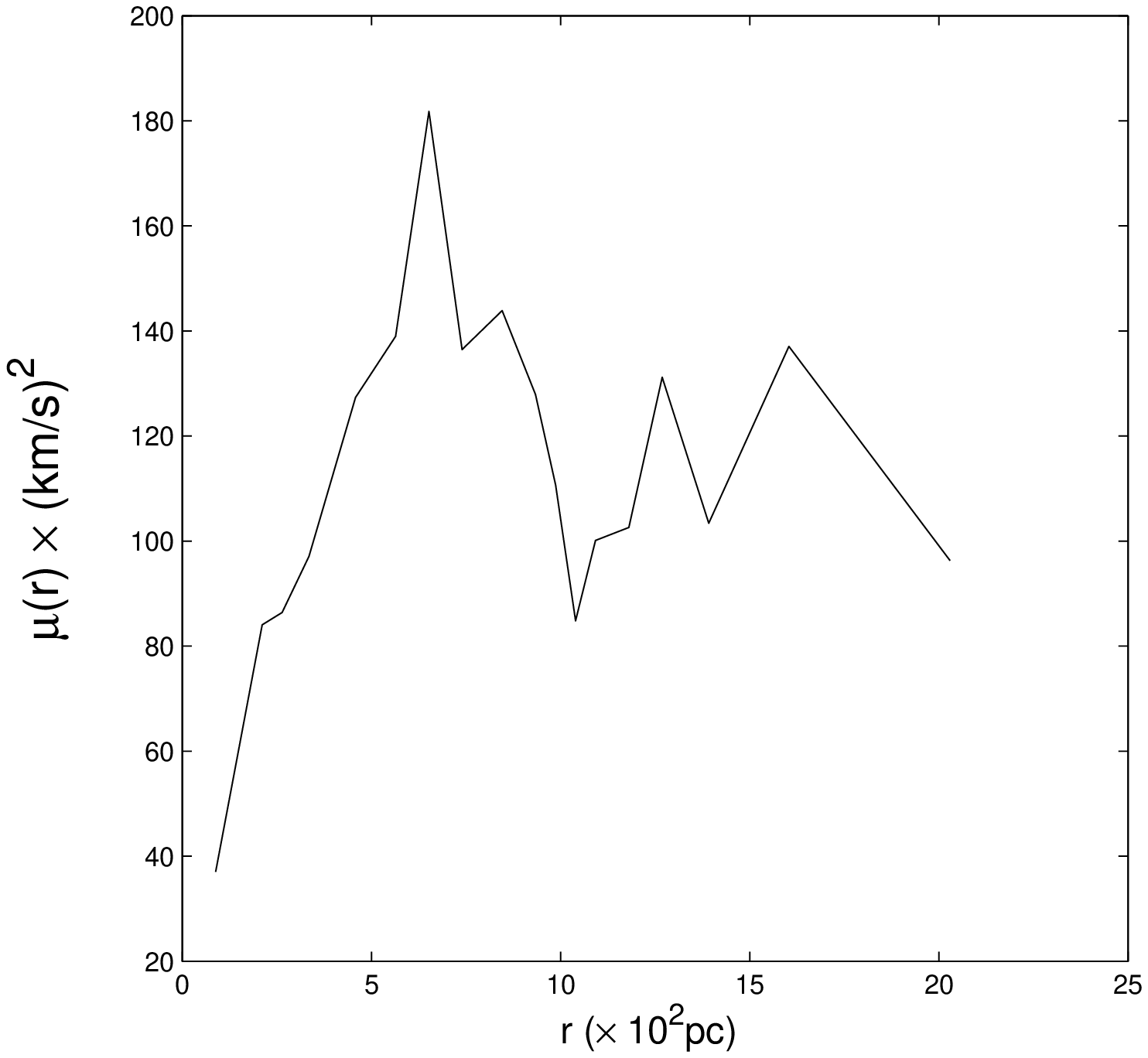}}
\end{center}
%\end{minipage}
%\begin{minipage}[c]{0.75\textwidth}
%\begin{center}
%\resizebox{3in}{3in} {\includegraphics{mu-fornax.eps}}
%\end{center}
%\end{minipage}
\end{center}\caption{\label{fig:fornax22} \small The mass distribution estimator for Fornax,
  $\hat{M}(r)$;  The radial velocity dispersion profile estimator,
  $\hat{\mu}(r)$ of stars }
\end{figure}

\subsection{Application of the Method:  The Fornax Dwarf Spheroidal Galaxy}

As an application to real observational data, we have use the
radial velocity data from Walker et al (2005) for the Fornax dSph
galaxy.  This dataset consists of velocities for 181 stars observed
one at a time over the past 12 years (Mateo et al. 1991 has a
description of some of the earliest data used here).  Data from newer
multi-object instruments are not included here (see Walker et al 2005b
for a first analysis of these newer results).  Full details about the
observations can be found in the papers referenced above; for the
present application we simply adopt these results along with their
quoted errors to see what sort of estimator for $M(r)$ we can extract
from these data.  From the simulations we already know that the
dataset is relatively small for this method, but these simulations
also imply that we have no reason to expect a strong bias in our mass
estimator due to the sample size, only a potentially large uncertainty
(at least a factor of 2) in the final mass estimate.

The histogram of the star counts from (Irwin and Hatzidimitriou 1995) is shown in the first panel of Figure
\ref{fig:fornax1}. The remaining two panels in Figure \ref{fig:fornax1} show the derived estimates for the
spatial density of stars in Fornax $\hat{f}(r)$ and the true projected radial distribution of stars in Fornax
$\hat{g}_{\bf Y}(y)$. The bottom panels of Figure \ref{fig:fornax21} give the estimated projected radial
distribution of stars from (Walker et al, 2004) and the estimated selection function $\omega(y)$ derived from
the data. Figure \ref{fig:fornax22} show the mass profile, $\hat{M}(r)$ and the squared velocity dispersion
profile, $\hat{\mu}(r)$.

Previous estimates of the Fornax mass have been based primarily on the classical, or 'King' (1966) analysis
outlined in the introduction of this paper, and give $M_{Fornax}$ range from $6$ to $10 \times 10^7$ M$_{\odot}$
(Mateo et al. 1991; Walker et al. 2005a). These models resemble truncated isothermal spheres and implicitly
assume spherical symmetry, velocity isotropy, that mass follows light, and a specific parametric form for the
joint density function $f_0(r,v)$. Our mass distribution estimator provides an independent measurement employing
neither of the latter two assumptions.  If one believes the second plateau seen at $r \ge 1.5$ kpc in Figure
\ref{fig:fornax22} (first panel) is a real feature, this would indicate a Fornax mass of $3.5 \times 10^8
M_{\odot}$.  However, we note that the simulations using Plummer models (Figure \ref{fig:diffsample}) show
similar plateaux at large radii related not to the underlying mass, but rather to the scarcity of remaining data
points.  The first plateau ($0.8 < r < 1.3$ kpc), by contrast, covers a region for which the Plummer simulations
indicate even a relatively small dataset can yield a reasonable estimate of the mass.  We therefore consider the
two plateaux in the $M(r)$ curve of Figure \ref{fig:fornax22} to bracket the region of plausible Fornax mass as
measured from an N $\sim$ 200 sample by the estimation technique.  The resulting interval of $1.0 \times 10^8
$M$_{\odot}$ $< $M$_{Fornax} $ $<$ $3.5 \times 10^8$ M$_{\odot}$ is of considerably larger mass than the results
of the classical analysis.

The estimation
approach then offers an attractive alternative to variants of the classical
analysis that require parametric dynamical models to interpret
kinematic data for dSphs.  Our simulations show that
this new non-parametric analysis will provide significantly more
accurate results as kinematic samples grow larger.

Finally, we can make some comments regarding the effect of velocity
anisotropy in our results.  If we consider an extreme case where the
outermost bins of the Fornax velocity dispersion profile are dominated
by stars in tangential orbits (that is, $\overline{{v_\theta}^2} >>
\overline{{v_r}^2}$; see Equations (\ref{eq:mrjeans}) and
(\ref{eq:jbeta})), then the outer bins give us the mass directly (from
Equation \ref{eq:mrjeans}) as $M(r) \sim r \overline{{v_\theta}^2}/G$,
or $10^8$ M$_{\odot}$, which is equivalent to the lower limit provided
by the mass estimation technique.  Thus, even if we consider an extreme
breakdown of the isotropy assumption, the data alone support a larger mass
than the simple King analysis.  Since it is more likely that the
velocity distribution has intermediate anisotropy (Kleyna et al.
2002), we conclude that the the mass of Fornax to the outermost
measured data point is between $1.0$ and $3.5 \times 10^8$ M$_\odot$
For an observed luminosity of $1.5 \times 10^7$ L$_{\odot}$ for
Fornax, this gives a global M/L ratio ranging from 7 to 22.  Thus,
even the most luminous, most baryonic-dominated dSph satellite of the
Milky Way is dominated by dark matter.

\section{Discussion}
We have presented a new method for estimating the distribution of mass in a spherical galaxy.  Along the way, we
have made some choices, generally preferring the simplest of several alternatives--for example, the use of
quadratic splines in Section 4 and simple histograms in Section 3.  In spite of this, the algorithm is a bit
complicated.  Is it really better than simpler methods--for example, using kernel smoothing to estimate the
distribution of projected radii and line of sight velocities and then using inversion to estimate $M(r)$?  Our
method differs from the simpler approach through its essential use of shape restrictions.  From a purely
statistical viewpoint, the monotonicity of $r^3\Psi''(r^2)/f(r)$ in Equation (\ref{eq:jns2}) provides a lot of
useful information [Robertson, et.al. (1988)].  Early in our investigations, we tried the kernel smoothing
outlined above and found that the resulting estimators did not satisfy the shape restriction, leading to
negative estimates of the mass density.  Imposing the shape restrictions directly guarantees non-negative
estimates, at least.  It also reduces the importance of tuning parameters, like the bandwidth in kernel
estimation or the number of bins in our work.  Imposing the shape restrictions also complicates the algorithm
and leaves a quadratic programming problem at the end.  We think that the game is worth the candle.

There is some similarity between our method and that of Merritt and Saha (1993) in that both use basis
functions, splines in our case, and both impose shape restrictions, the condition that $f
> 0$ in theirs.  Our method makes more esssential use of shape restrictions and treats the inversion problem in
greater detail.

\vskip2em

This research was supported by grants from the National Science
Foundation and the Horace Rackham Graduate School of the University of
Michigan.

%Similarly, the model (\ref{eq:spln1}) for $M(r)$ has the
% wrong asymptotic behavior as $r$ approaches $0$.  Alternative models
% include higher order splines,
% $$
%         M(r) = \sum_{i=1}^m \beta_i(r-r_{i-1})_{+}^{p}
%         $$
%         with $p \ge 2$.  For $p = 3$, $M(r) \sim \beta_1r^{3}$ as
%         $r \to 0$, and $3 \beta_1/4\pi$ provides an estimate of the
%         central mass density.  Unfortunately, translating the shape
%         restrictions into conditions on $\beta_1,\cdots,\beta_m$ is
%         much more difficult when $p = 3$ then when $p = 1$.
% $$
%         \cdots
% $$

\end{document}